\documentclass[onecolumn]{aastex6}
\usepackage{mathrsfs}
\usepackage{amsmath}

\begin{document}

\title{Introduction to Magnetic Reconnection}

\author{Amir Jafari}

\and

\author{Ethan Vishniac}
\affil{Department of Physics \& Astronomy, Johns Hopkins University, Baltimore, MD, USA}

\begin{abstract}

We review the basic concepts of magnetic reconnection and propose a general framework for the astrophysical reconnection at large scales. Magnetic reconnection is the rearrangement of magnetic field topology. The conventional Sweet-Parker scheme and some of its extensions presume a non-turbulent medium and generally produce very slow reconnection or an unstable configuration. However, the assumption of laminar flow is unrealistic in astrophysics since, even in an initially quiet environment, magnetic reconnection by itself can drive turbulence. The resulting turbulence has the potential to enhance the reconnection rate. This can lead to an unstable feedback loop as reconnection drives turbulence and turbulence drives reconnection. Stochastic reconnection was proposed, and subsequently tested by numerical simulations, for high $\beta$ plasmas with a magnetic Prandtl number of order unity, $Pr_m\sim 1$. This model predicts reconnection speeds comparable to the large scale turbulent eddy velocity. A recent study of stochastic reconnection for $Pr_m>1$ has shown that the width of the outflow layer and the ejection velocity of matter from the reconnection region seem to be unaffected by viscosity in typical astrophysical systems. However if $Pr_m>1$ viscosity can suppress small scale reconnection events near and below the Kolmogorov or viscous damping scale. This will produce a threshold for the suppression of large scale reconnection by viscosity when $Pr_m$ is larger than the root of the Reynolds number, $Pr_m>\sqrt{Re}$. For $Pr_m>1$ this leads to the spectral index $\sim-4/3$ for length scales between the viscous dissipation scale and eddies larger by roughly $Pr_m^{3/2}$. 
\end{abstract}

\section{Introduction}\label{s1}

\begin{figure}
\includegraphics[scale=.43]{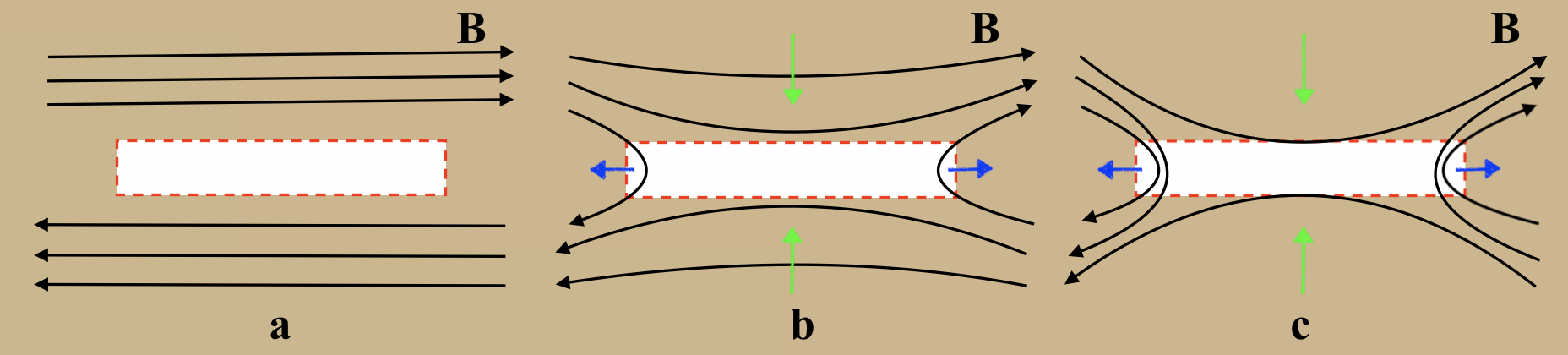}
\centering
\caption {\footnotesize {Conventional reconnection: initially antiparallel magnetic field lines in a conductive medium approach in a reconnection zone with locally strong field gradients.}}
\label{general-reconnection}
\end{figure}

One way to lower the magnetic energy stored in a low-$\beta$ plasma is to reorganize the magnetic field lines into configurations with lower energies. The tendency of physical systems to achieve the lower energy states available implies that the reconnecting magnetic field lines to arrive at a lower energy state should be a natural tendency in conductive fluids. The concept of magnetic reconnection was introduced in the late 1950s to explain the efficient transformation of magnetic energy to kinetic energy observed in the solar flares and also interaction between Earth's magnetosphere and interplanetary medium (Parker 1970; Innes et al. 1997). When magnetic reconnection occurs, constraints of an ideal fluid is broken in a relatively small region that affects large scale structures of the plasma and allows efficient transformation of magnetic to kinetic energy. When two plasmas, which are initially separate, come into contact with one another, for instance when solar wind interacts with the Earth's magnetic field, they cannot mix. A thin boundary layer forms between the two plasmas separating their magnetic fields; Fig.(\ref{general-reconnection}). In equilibrium, pressure balance will determine the location of this boundary layer. On either side of this boundary layer, magnetic fields have different strengths and orientations and thus the boundary layer constitutes a current sheet which is also called a diffusion layer or resistivity layer reflecting the idea that the magnetic field is changing over a finite, i.e. nonzero, distance. The physics in this layer is dominated by resistive diffusion of the magnetic field (see e.g., Zweibel \& Yamada 2009; Yamada et al. 2010). It should be emphasized that in fact Ohmic dissipation by itself converts only small amounts of magnetic energy to heat. The major energy conversion mechanism results from Ohmic dissipation acting differentially to change the magnetic field topology such that they lead to a strong curvature in the current sheet. This change in the field topology releases small amounts of energy, however, the curved force lines, unfolding and accelerating the plasma out of the current sheet, accelerate the plasma and increase the kinetic energy by annihilating the magnetic energy (see recent reviews by Yamada et al. 2016 and Loureiro \& Uzdensky 2016).

The interaction of the field lines leads to the formation of a singular current sheet in which the parallel electric field becomes very strong and consequently the field lines lose their identity (Yamada et al. 2010). In fact, it was suggested long time ago (Dungey 1954) that the collapse of the magnetic field in an X-type magnetically neutral point can lead to the formation of such current sheets which eventually leads to reconnection; Fig.(\ref{general-reconnection}). During reconnection, the magnetic field topology changes and $\bf{j\times B}$ forces lead to the conversion of magnetic energy to particle energy. The structure of the current sheets in symmetric situations has been studied extensively in the last several decades (see Waelbroeck 1989; Becker et al. 2001; Jemella et al., 2004). However, as pointed out by Yamada et al. (2010), many realistic situations such as the solar flares are far from being symmetric, so there is no reason to believe that current layers are as long as the global size of the system. Parker (1979) demonstrated the inevitability of the 3D current sheet formation with  lengths comparable tp the local scales over which the ambient magnetic field changes by a finite amount. A general expectation is that in highly collisional plasmas, a rectangularly shaped Sweet-Parker current sheet forms. On the other hand, for collisionless plasma, the shape of the current sheet would be of the Petschek-type X-point double wedge with a much faster reconnection rate (Yamada et al. 2010).

Reconnection is indeed a ubiquitous process in astrophysical environments where two plasmas with opposite magnetic flux regions are brought into contact. As a matter of fact, fast reconnection, of order of the local Alfv\'en speed, is a crucial requirement for the astrophysical dynamos where the field needs to reconnect on dynamical, rather than resistive, time scales (Lovelace 1976; Priest \& Forbes 2002). Understanding reconnection is crucial to explain the $\gamma$-ray bursts (Lyutikov \& Lazarian 2013), solar wind and flares (Sturrock 1966) and stellar flares (Parker 1993; Biskamp 1993 and references therein). Nevertheless, it should be noted also that reconnection is not so easy to observe since it transfers most of the magnetic energy into kinetic motion of smaller eddies supporting the energy cascade (Lazarian et al. 2015). Early reconnection models (Parker 1957; Sweet 1958; Petschek 1964) employed magnetohydrodynamics (MHD) with the assumption that electrons and ions move together as a single fluid. Indeed, Parker (1957) and Sweet (1958) independently suggested the first model for magnetic reconnection. However, their simple model predicted a reconnection speed that was very slow compared with the observations (Kulsrud 2000; Biskamp 1997). Of course, the single-fluid picture has to be modified if applied in a thin reconnection region, e.g., those in the magnetosphere in which ions become demagnetized and the relative drift velocity between the electrons and ions can be large. Reconnection rate seems to increase by increasing the collision mean free path leading to the two-fluid regime (see Yamada et al. 2010 for the references). On the other hand, reconnection layers, such as those created at the magnetopause, may have thicknesses comparable to the ion skin depth. Because of the force balance between the magnetic field and the plasma kinetic pressure, the ion skin depth may also be of order the ion gyroradius (Yamada et al. 2010). This also breaks the notion of a single fluid and can lead to a large electric field at the reconnection region speeding up the reconnection rate ( see e.g., Vasyliunas 1975; Dungey, 1995; Kivelson \& Russell 1995).

By the 1980, the progress in numerical computations made it possible to test different theories on reconnection by MHD simulations (Sonnerup 1970; Vasyliunas 1975; Ugai \& Tsuda 1977; Hayashi \& Sato 1978; Priest \& Forbes 1986; Biskamp 1986). The overall result was disappointing; no match with the available theories. Rapid improvements in numerical computations since 1990s have led to the possibility of applying realistic boundary conditions in 3D, modeling current sheet instabilities and treating ion and electron fluids separately by adding terms like electron pressure (second Hall term) to the Ohm's law. Fluid models require very frequent collisions between ions and electrons which is an absent feature in collisionless plasmas where fast reconnection is observed. In order to account for the kinetic effects, we need to go beyond fluid models by, for example, adding correction terms to fluid models (see, e.g., Kuznetsova et al., 2007). Kinetic effects beyond two-fluid models include acceleration or heating of charged particles, non-gyrotropic pressure, and instabilities due to inhomogeneities in velocity space (Yamada et al. 2010). 
 
  \begin{figure}
\includegraphics[scale=.5]{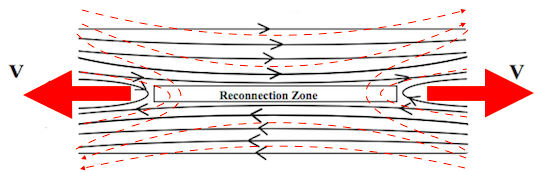}
\includegraphics[scale=.3]{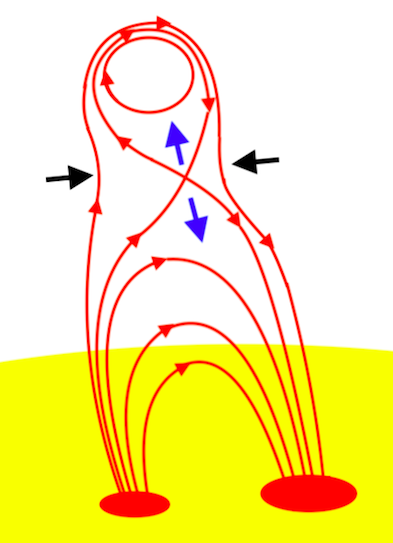}
\centering
\caption {\footnotesize {Left: The Sweet-Parker model. A reconnection zone forms between two approaching regions of plasma with oppositely oriented magnetic fields. The reconnected field lines are shown as red dashed lines. Right: A typical reconnection event on the solar surface.}}
\label{Sweet-Parker}
\end{figure} 
One naive feature of the Sweet-Parker model and its several extensions is the assumption of laminar flow. It is naive since reconnection itself, as well as different instabilities, has the potential to make the medium turbulent. Moreover, almost all flows in astrophysical media are turbulent to some degree. Other instabilities in a quiet medium, such as tearing modes and magnetorotational instabilities, can also produce or enhance the turbulence. Of course, we may distinguish between the external turbulence, initially and independently present in the environment, and the internal turbulence which is produced as a result of reconnection or other magnetohydrodynamic instabilities like tearing modes. In any case, the assumption of a laminar flow should be relaxed. After all, turbulence is ubiquitous in astrophysics and, more importantly, it can also enhance reconnection rates to the order of turbulent velocity (see e.g., the review by Lazarian et al. 2015). In fact, magnetic reconnection rate is enhanced by turbulence and, on the other hand, turbulence requires magnetic reconnection in order to evolve in a self-similar way (Lazarian et al. 2012). In short, despite the fact that a "direct" connection between turbulence and the reconnection rate has been questioned by some authors (see reviews by Guo et al. 2012; Matthaeus et al. 2015; Cranmer et al. 2015) but it is generally believed that turbulence affects, and most likely enhances, the reconnection rate (see e.g., Lazarian et al. 2015).

It might seem that the simplest way to take turbulence into account is through introducing an enhanced magnetic diffusivity---turbulent diffusivity. Since the Sweet-Parker reconnection rate depends on the molecular magnetic diffusivity, which is negligible in almost all astrophysical contexts, therefore an enhanced turbulent diffusivity might sound like a promising approach to somehow higher reconnection rates. Nevertheless, the substitution of microscopic diffusion by a turbulent transport coefficient needs to be justified on theoretical grounds (Vishniac \& Lazarian 1999). Some dynamo theories apply the notion of an enhanced turbulent diffusivity in order to avoid the flux-freezing conditions associated with the high electric conductivity of astrophysical plasmas. With flux-freezing, the topology of magnetic field cannot change and consequently small scale loops of magnetic field in a turbulent flow cannot generate large scale fields. Unfortunately, this generally leads to an ill-founded theory since strong fields should prevent the turbulent mixing of oppositely oriented fields (Parker 1992; Lazarian and Vishniac 1999 henceforth LV99). This also leads to a failure of kinematic dynamo theory (Cattaneo \& Vainshtein 1991, Kulsrud \& Anderson 1992). 
\iffalse

\begin{figure}
\includegraphics[scale=.6]{current-sheet}
\centering
\caption {\footnotesize {Formation of current sheet by externally driven flow (Forbes 2007)}}
\label{current-sheet}
\end{figure} 

\fi

On the other hand, even in the absence of any large scale current sheet, where reconnection occurs over scales of order several thousand ion gyroradii, the magnetic field can reconnect in much smaller scales in the turbulent cascade. In fact, energetic intermittent reconnection events should be ubiquitous in the small scales as the turbulent field component is drifted with the fluid. In other words, Richardson advection has the potential to bring the initially distant field lines to small separations of order the ion gyroradii which in turn leads to multiple spontaneous reconnection events (Eyink et al. 2013). Lazarian and Vishniac (LV99) proposed a model of fast reconnection in a turbulent and magnetized plasma by identifying stochastic wandering of the magnetic field lines. This model, initially proposed for totally ionized and inviscid plasmas, was later shown to be valid also for partially ionized collisionless plasmas up to a certain percentage of neutrals and also for the collisional gases with the magnetic Prandtl numbers less than unity $Pr_m< 1$ (Lazarian et al. 2004). The underlying mechanism suggested by LV99 yields very fast reconnection in a good agreement with the numerical simulations (Kowal et al. 2009). Stochastic reconnection, as employed in LV99, results from the magnetic field line wandering in which the reconnection rate does not depend on the details of current sheet structure. In fact, it is given by the ability of ejected plasma to diffuse away along the magnetic field lines (Lazarian et al. 2015). Eyink et al. (2011) showed that the detailed predictions of LV99 also follow from the stochastic flux-freezing laws. LV99 treated stochastic reconnection as a process similar to the Sweet-Parker model but with the vital difference that the stochastic field line wandering leads to a much broader outflow region. Also, the properties of this outflow region do not depend on neither the width of the current sheet nor the value of the Lundquist number. Numerical simulations testing LV99 (Kowal et al. 2009) have confirmed the Sweet-Parker scaling in the absence of turbulence and LV99 scaling in the presence of turbulence. Nevertheless, this model is not appropriate for reconnection processes at the scale of ion Larmor radius, for example, in the Earth magnetosphere (Lazarian et al. 2015).

  \begin{figure}
\includegraphics[scale=.5]{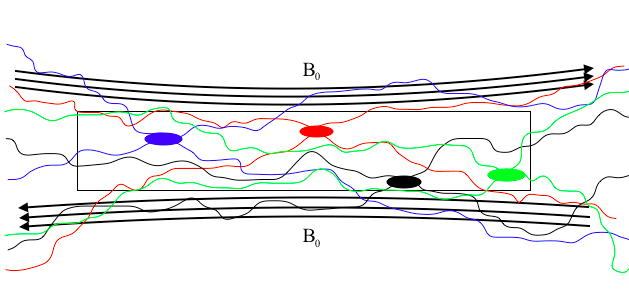}
\centering
\caption {\footnotesize {In a turbulent medium, in addition to the large scale field $B_0$ there is a also small scale magnetic field. The small scale magnetic field lines wander around stochastically and everywhere that two field lines come to a close contact, a local Sweet-Parker reconnection occurs. Here, only four of many simultaneous reconnection events have been shown.}}
\label{ST1}
\end{figure}

In many cases in astrophysical turbulent media, the fluid is not inviscid and in fact the viscosity is much larger than the resistivity. In the latter case, known as the high magnetic Prandtl number regime, the substructure of the magnetic field would not continue smoothly down to resistive scales without this important assumption that the resistivity is roughly equal to the viscosity. LV99 is not applicable in this high magnetic Prandtl number regime; $Pr_m={\nu}/{\eta} \gg 1$ where $\nu$ is kinematic viscosity and $\eta$ is magnetic diffusivity. One example is the interstellar medium (ISM) which is also collisional and turbulent. In star formation processes, as well, the partially ionized gas has a viscosity much larger than the resistivity. The same regime is encountered also in galaxies, protogalaxies and clusters filled with hot and rarefied plasmas. In an effort to construct a theoretical model for magnetic reconnection in such environments, Lazarian et al. (2004) aimed to extend the model of LV99 for the case of a magnetized and partially ionized medium. The authors considered the effect of a large neutral fraction on a strongly turbulent cascade in a magnetized plasma and applied this to the problem of magnetic reconnection in a partially ionized gas. Numerical simulations, however, did not support their current sheet structure. Jafari et al. (2018) developed an analytical model of stochastic magnetic reconnection for high magnetic Prandtl numbers and suggested a general model of stochastic reconnection for both collisional and collisionless regimes with arbitrary degree of ionization.

In situ measurements of reconnection events, e.g., in the Earth's magnetosphere, provide primary data to model reconnection on small scales namely of order the ion inertial length. These scales, however, are not of interest in the astrophysical applications where the reconnection relies on large-scale MHD turbulence rather than small scale plasma physics (Lazarian et al. 2015). In fact, there is a spectrum of different scales over which reconnection may occur. This situation resembles that in quantum field theories where we compensate our ignorance of very small scale physics by renormalization. Similarly, to avoid the complexity appearing in length scales smaller than a cut-off, e.g., the ion inertial length where the two-fluid effects become important, we may model the overall effect by resorting to the r.m.s separation that magnetic field lines experience. The fact that reconnection results in a change in the magnetic field topology is suggestive of considering the reconnection employing a diffusion process. In the normal diffusion, the mean square distance between particles grows linearly with time as $y^2(t)\sim t$. This is a random walk process and can be applied to the magnetic field lines as well. The classical Sweet-Parker model, historically constructed in a completely different way, is recovered by choosing the magnetic diffusivity as the diffusion coefficient. Replacing it with others, e.g., the Bohm diffusion coefficient, turbulent diffusivity and anomalous diffusivity to name a few, may enhance the diffusion and so the reconnection rate a little bit depending on the prevalent conditions. However, these conditions may also enforce a completely different mechanism of diffusion so a mere replacement of diffusion coefficient would not always work. For instance, turbulence leads to a super-diffusion, known as Richardson diffusion, in which the mean square distance grows non-linearly with time as $y^2(t)\sim t^3$. How can we unify all these apparently different schemes of reconnection in a general framework? We will construct such a theoretical framework---a standard model--- without developing a new theory for reconnection itself by assuming that astrophysical reconnection in large scale systems relies on a diffusion process.

Despite all the theoretical and numerical efforts to understand reconnection, there are many questions yet to be answered. How exactly magnetic energy is converted to the particle energy; how the current sheet instabilities affect the reconnection process; how magnetic reconnection can be so fast as observations indicate and many other questions are few examples of such questions. Extensive reviews have discussed these aspects of reconnection (see e.g., Zweibel \& Yamada 2009; Yamada et al. 2010; Lazarian et al. 2015; Zweibel \& Yamada 2016). Here, we will be mostly concerned with the problem of the fast reconnection in astrophysical contexts. After a short review of some conventional models in \S{2}, we turn to discuss the importance of turbulence in \S {3}. In \S 4, we show how different models of reconnection can be unified in a general formalism of three dimensional diffusion. In \S {5}, we focus on stochastic reconnection and show how its different regimes emerge from the general formalism introduced in \S {4}. The final conclusions are discussed in \S{6}.

\section{Conventional Models}\label{s2}

In a plasma with magnetic diffusivity $\eta$ and velocity field $\bf{v}$, the induction equation for the magnetic field $\bf{B}$ reads

\begin{equation}\label{6}
\frac{\partial {\bf{B}}}{\partial t}=\eta \nabla^2 {\bf{B}}+\nabla \times ( \bf{v} \times \bf{B}).
\end{equation}
In eq.(\ref{6}), the first term in the RHS represents diffusion whereas the second term represents the convection. If the system has the typical length scale $\textit{L}$, then we have

\begin{equation*}
\eta \nabla^2 \textbf{B} \sim \frac{\eta \textit{B}}{L^2}, \; \nabla\times (\textbf{v} \times \textbf{B}) \sim \frac{\textit{vB}}{L},
\end{equation*}
 
With a small magnetic Reynolds number $R_m=Lv/\eta$, the convective term is negligible with respect to the diffusive term and thus we have ${\partial \textbf{B}}/{\partial t}=\eta {{\nabla}}^2 \textbf{B}$. The latter diffusion equation implies that the field variations on a length scale $L$ are decayed over the diffusion time scale, $\tau_R=L^2/\eta$. For the large magnetic Reynolds numbers, $R_m \gg 1$, the diffusion term is negligible and magnetic field lines move with the plasma, ${\partial\textbf{B}}/{\partial t}={\nabla} \times (\textbf{v} \times \textbf{B})$. This is the ideal case of the evolution of magnetic field, in which, the force lines are attached or frozen-in into the plasma. Also note that one may rewrite the last equation as

\begin{equation*}
\frac{\partial\textbf{B}}{\partial t}=(\textbf{B.}{\nabla})\textbf{v}-\textbf{B}({\nabla} \textbf{.v}).
\end{equation*}

The first term in the above equation shows that any accelerating motion along the magnetic field lines will amplify the field, and also any shearing motion perpendicular to the field lines will make a change in magnetic field direction by increasing the field component along the direction of the flow. The second term implies that an expansion in the flow, when $\nabla \textbf{.v}>0$, decreases the field whereas a compression, when $\nabla \textbf{.v} <0$, amplifies the magnetic field. Integrating eq.(\ref{6}), it can be shown that the magnetic flux is conserved through any closed curve that moves with the plasma when $\eta\rightarrow 0$. This result, called the Alfv\'en frozen flux theorem, means that the flux through a comoving surface $\textit{C}$, moving with the plasma velocity, is conserved. Consequently, the plasma can move along with the magnetic field lines and for its perpendicular motion with respect to the force lines either the field lines push the plasma, or they are dragged with it. Yet another way, with more mathematical rigor, to look at this problem is to write the divergence-free magnetic field using the Clebsch coordinates (Yamada et al. 2010). We write ${\bf{B}}=\nabla\alpha\times\nabla\beta$ where $\alpha$ and $\beta$ are two scalar functions of time and space such that
\begin{equation}
{d\alpha\over dt}={\partial \alpha\over\partial t}+{\bf{v.}}\nabla\alpha=0.
\end{equation} 
\begin{equation}
{d\beta\over dt}={\partial \beta\over\partial t}+{\bf{v.}}\nabla\beta=0.
\end{equation} 

The force lines are defined by time independent functions $\alpha=const.$ and $\beta=const.$ which show the field is moving along with the matter implying flux freezing. The topology of the magnetic field is the major player in determining the static plasma equilibrium. If we vary the magnetic energy density 

\begin{equation}{\cal{L}}_m={B^2\over 2}={1\over 2} (\nabla\alpha\times\nabla\beta)^2,
\end{equation}

with respect to $\alpha$ and $\beta$, we find $\bf{j\times B}=0$ which is the condition of a force-free equilibrium (Yamada et al. 2010).

We also note in passing that even with a non-zero resistivity, it might be possible to obtain a similar result by defining the magnetic flux transporting velocity $\textbf{u}$ using eq.(\ref{6}),

\begin{eqnarray}\label{10}
{\nabla} \times (\textbf{u} \times \textbf{B})&=&\frac{\partial \textbf{B}}{\partial t} \\
&=&\eta {{\nabla}}^2 \textbf{B}+{\nabla} \times (\textbf{v} \times \textbf{B}) \notag ,
\end{eqnarray}

in which instead of plasma velocity, we used the flux velocity, $\textbf{u}$. However, this raises the question of the existence and uniqueness of the flux transporting velocity. To obtain a sufficient condition for the existence of this velocity, we can integrate eq. (\ref{10}) which yields $\textbf{E}+\textbf{u} \times \textbf{B}={\nabla} F$ where $F$ is an arbitrary function. If we compare this equation with the Ohm's law, we can write $\eta{\bf{j}}=\boldsymbol{\nabla}F+(\textbf{v}-\textbf{u}) \times \textbf{B}$. If $\textbf{B}\neq 0$, a sufficient condition for the existence of the flux transporting velocity, $\textbf{w}$, can be obtained, $\textbf{B.}{\nabla}F=\eta{\bf{j.B}}=\textbf{B.}\textbf{E}$. The latter equation can be solved, for example, if $\textbf{J}$ is perpendicular to $\textbf{B}$ or $\textbf{E.B}=0$, therefore $F=0$ is a trivial solution in this case. Here, our aim is not to study the procedure of solving the above equation, however it is worth to mention that even when $\textbf{B.J}\propto\textbf{B.E} \neq 0$ one can still find $\bf{F}$ assuming that there is a "transversal surface " such that the force lines cross it just once (Wilmot-Smith et al. 2005).

\subsection{Resistive Diffusion: Sweet-Parker Model}\label{ssSP}
In an electrically conducting medium, plasma pressure stretches out the large scale flux tubes, in which by assumption the plasma is trapped, while they cannot release their huge amounts of energy to return to a relaxed state. Consequently, there would be no efficient transformation of magnetic energy to kinetic energy in highly conductive plasmas (Biskamp 2003). With a finite resistivity, electric current is dissipated by the Ohmic dissipation especially in the current sheet where an intensive electric current flows. This by itself dissipates the magnetic energy a little bit not too much, however, it leads to a sudden change in the field topology---reconnection---that transfers magnetic energy to particle energy allowing the plasma to relax to a lower energy state.

  \begin{figure}
\includegraphics[scale=.45]{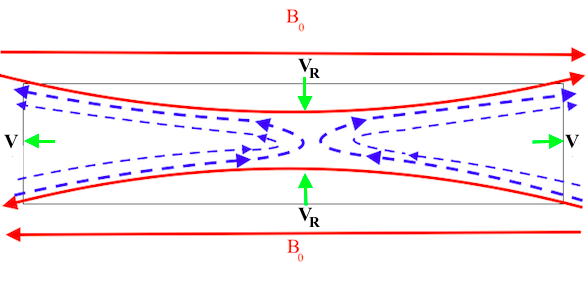}
\centering
\caption {\footnotesize {Sweet-Parker reconnection, in a non-turbulent medium with magnetic diffusivity $\eta$ and Lundquist number $S$, can be understood by considering the normal diffusion of the field lines as indicated by induction equation $\partial_t{\bf{B}}=\eta\nabla^2{\bf{B}}$. Diffusion of the field lines over the length of the current sheet $L_x$ leads to a "field line separation" $y^2\sim\eta t\sim \eta L_x/V_A$ where $V_A$ is the Alfv\'en speed. Mass conservation, $L_x V_R=y V_A$ leads the reconnection speed $V_R\sim V_A/\sqrt{S}$. }}
\label{SP}
\end{figure} 

A schematic of a typical reconnection zone can be seen in Fig.(\ref{SP}): two regions with oppositely oriented magnetic fields are pushed toward each other which indicates an inflow of matter towards the intermediate reconnection layer with reconnection speed $V_R$. Matter is then ejected, in the horizontal direction, with a velocity of almost the local Alfv\'en speed $V_A$ in the magnetized region. In order to estimate the latter, one can assume that the magnetic energy $B^2/2$ is totally converted to the kinetic energy of the outflow which moves with velocity $v_x$;
\begin{equation}
\rho v_x^2\simeq B^2/2,
\end{equation}
where $\rho$ is the density. This leads to an ejection speed of order the local Alfv\'en speed, $v_x \simeq V_A$.

\iffalse

We can write the conservation of energy in a reconnection zone of length $\Delta$ and width $\delta$. The Poynting energy flux into the zone is $V_R B^2\Delta$. This energy is consumed in two ways: Joule dissipation $J^2 \eta \delta \Delta $ and the kinetic energy flux of the outflow $\rho V_A^2 (V_A \delta)$ (Zweibel \& Yamada 2009). After some reshuffling, we find
\begin{equation}
V_A^3 \delta=V_{R} V_A^2\Delta-\epsilon \Delta\delta,
\end{equation}
where $\epsilon={\bf{E.J}}/\rho$ is the energy dissipation rate. Neglecting the dissipation, the last term, we would recover the mass conservation $V_A \delta=V_{R} \Delta$.

\fi

In order to estimate the inflow or reconnection speed, which has been the subject of many works since early 1960s due to unrealistic results, we start with the Sweet-Parker scaling. In a plasma with a finite diffusivity $\eta$, the Ohm's law reads 
\begin{equation}\label{Ohm1}
 {\bf{E}}+{\bf{v\times B}}=\eta {\bf{J}},
 \end{equation}
which leads to $J \sim  {V_RB}/{\eta}$ where $V_R$ stands for the inflow or reconnection velocity. The physical content of the above relation for the electrons is that, in the steady state, the Lorentz force and the collisional drag force balance. On the other hand, for a current sheet of thickness $ \delta$ and length $\Delta$, in the steady state, one can apply the Ampere's law to estimate the current, $J \sim {B}/{\delta}$ and thus we get $V_R\sim {\eta}/{\delta}$. 

The last piece comes form mass conservation. We can write the conservation of energy in a reconnection zone of length $\Delta$ and width $\delta$. The Poynting energy flux into the zone is $V_R B^2\Delta$. This energy is consumed in two ways: Joule dissipation $J^2 \eta \delta \Delta $ and the kinetic energy flux of the outflow $\rho V_A^2 (V_A \delta)$ (Zweibel \& Yamada 2009). After some reshuffling, we find
\begin{equation}
V_A^3 \delta=V_{R} V_A^2\Delta-\epsilon \Delta\delta,
\end{equation}
where $\epsilon={\bf{E.J}}/\rho$ is the energy dissipation rate. Neglecting the dissipation, the last term, we would recover the mass conservation $V_A \delta=V_{R} \Delta$. 

Putting all this together, we obtain a reconnection speed of order

\begin{equation} \label{15}
V_R=\left  ( {\eta \frac{V_A}{\Delta}}\right)^{1/2}=V_AS^{-1/2},
\end{equation}

 where $S=V_A\Delta/ \eta$ is the Lundquist number. Note that the Sweet-Parker time scale $t_R=\delta\Delta/\eta$ is shorter than the resistive time scale $t_\eta=\Delta^2/\eta$ by a factor of $\sqrt{S}$; $t_R=t_\eta/\sqrt{S}$ and longer than the Alfv\'en time scale $t_A=\Delta/V_A$ by the same factor; $t_R=\sqrt{S} t_A$.
 
\begin{figure}
\includegraphics[scale=.65]{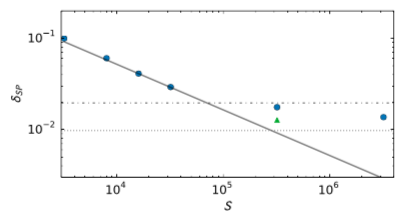}
\centering
\caption {\footnotesize {Current sheet width $\delta_{SP}$ in units of box size of the current sheet during quiescent reconnection prior to the onset of instability, from numerical simulations performed by Oishi et al. (2015). The circles show simulations with varying Lundquist number $S$, the solid line gives the Sweet-Parker scaling, and the triangle shows a run at double resolution. The dotted-dashed line shows a resolution of $10$ zones for standard ($512^3$-equivalent) runs, while the dotted line shows $10$ zones for the high resolution ($1024^3$- equivalent) run. }}
\label{Oishi2015}
\end{figure}

 The Sweet-Parker scheme can also be understood in terms of magnetic field diffusion. The average distance between a pair of field lines $\delta(t)$ increases with time $t$ as $\delta\simeq \sqrt{\eta t}$. This is normal diffusion, in which average square distance scales linearly with time; see Fig.(\ref{Oishi2015}). Substitution of the Alfv\'en time $t_A=\Delta/V_A$, and using mass conservation, we recover eq.(\ref{15}) above.

In the solar corona, where $S$ is of order $10^{12}$, the above expression leads to a reconnection time of order $t_R \geq 10^6\;s$. However, the measured time scale is of order $t_R \sim 100\;s$. For instance, the field topology in the soft-x-ray pictures changes in a time scale of minutes or at most hours which is much shorter than the Sweet- Parker time (Parker 1957). Thus, in spite of the fact that the Sweet-Parker scheme predicts much faster conversion rate for magnetic energy than the global diffusion, nevertheless, it is still much too slow compared with the observations (Yamada et al. 2010). Also note that with vanishing resistivity, the width of the current sheet vanishes as well, and reconnection may only proceed with an anomalous resistivity discussed below (see also Shay et al. 2001).

\subsection{Bohm Diffusion}
The magnetic diffusivity used in the above considerations of the Sweet-Parker scheme is thought to be a constant characteristic of the plasma. However, this picture may change, at least in theory, in several different situations. For example with the Bohm diffusion---diffusion of plasma across the magnetic field---the diffusivity should be replaced with the Bohm diffusion coefficient, $\eta_{B}$. To get an estimate of this coefficient, we note that the diffusion coefficient for collisions is generally given by
$D={\lambda^2}/{\tau} \simeq u^2\tau$ where $\lambda$ is the mean free path, $\tau$ is the average time between two collisions and $v$ is the average speed. Suppose that the collision frequency $\tau^{-1}$ is  larger than the gyrofrequency $\omega=eB/m$, $e$ and $m$ being respectively the electron's electric charge and mass. Hence the particle would experience many collisions long before completing its orbit across a magnetic field line. Thus it can be considered moving with thermal velocity $v_{th}$ which leads to the classical diffusion coefficient 
\begin{equation}
D_\tau=v_{th}^2\tau,
\end{equation}
where $v_{th}=\sqrt{k_B T/m}$ with Boltzmann constant $k_B$ and temperature $T$. On the other hand, if the collision frequency is smaller than the gyrofrequency, $\tau^{-1}\ll \omega$, the particle would have enough time to orbit around the field line between its two subsequent collisions so the diffusion coefficient becomes 
\begin{equation}
D_\omega={r^2\over\tau}={D_\tau\over(\omega\tau)^2},
\end{equation} 
where we have used $r=v_{th}/\omega$. The maximum diffusion happens for $D_\omega=D_\tau$ that is when the collision frequency equals the gyrofrequency, $\omega\tau=1$, and we get the Bohm diffusion:
\begin{equation}
D_B \simeq {k_B T/eB}.
\end{equation}
Thus, the Bohm diffusion scales as $1/B$ in terms of the magnetic field. In a magnetized plasma, the collision frequency is usually smaller than the gyrofrequency thus the classic diffusion coefficient, $D_\omega=r^2/\tau$, is proportional to $1/B^2$ that is $D_\omega \propto \omega^{-2}\propto B^{-2}$.

For the diffusion of magnetic field lines, instead of the particle thermal velocity $v_{th}$, one should use the Alfv\'en velocity $V_A=B/\sqrt{4\pi\rho}$:

\begin{equation}\label{25}
\eta_{B} \simeq V^2_A \left(\frac{m}{eB}\right)=r V_A ,
\end{equation}

where we have used $r=mV_A/(eB)$ as the cyclotron radius. Substituting $\eta_B$ in the Sweet-Parker reconnection speed yields $V_B=\sqrt{\eta_B V_A / \Delta}$ (see e.g., Parker 1979; Lazarian \& Vishniac 1999), we find the Bohm reconnection speed as

\begin{equation}\label{26}
V_{B}=V_A \left (\frac{r}{\Delta}\right)^{1/2}.
\end{equation}

This result shows that the Bohm diffusion can produce fast reconnection when $\Delta \sim r$. Nevertheless, this is hardly satisfied in real situations specially in astrophysical environments. The implication is that the Bohm diffusion would not generally lead to fast reconnection rates.

\begin{figure}
\includegraphics[scale=.65]{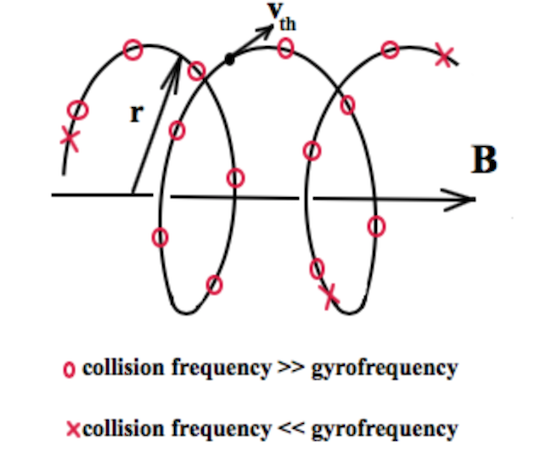}
\centering
\caption {\footnotesize {The orbit of a particle with gyroradius $r$ and thermal velocity $v_{th}$ in a magnetic field $\bf{B}$. Small circles and crosses indicate possible collision points with other particles in different regimes. In the highly collisional regime, in which the collision frequency $\tau^{-1}$ is much larger than the gyrofrequency $\omega=eB/m$, the particle has a very short time to feel the magnetic field so it can be considered moving with thermal velocity $v_{th}$ which leads to the diffusion coefficient $D=v_{th}^2\tau$. With a collision frequency much smaller than the gyrofrequency, $\tau^{-1}\ll \omega$, the particle has enough time between two collisions to complete its orbit around a field line so the diffusion coefficient becomes $D=r^2/\tau$. }}
\label{Bohm}
\end{figure} 

\subsection{Hyper-resistivity}
Other diffusion coefficients, similar to the Bohm coefficient, have been also proposed to enhance the  resistivity in the Sweet-Parker model. However, it turns out, this may become something deeper than a simple replacement. The equation of motion for the charged particles can be reshuffled into a form similar to the Ohm's law. The equation of motion for the species $\alpha$ in a plasma, where $\alpha$ may represent ions $\alpha=i$, neutrals, $\alpha=n$ or electrons $\alpha=e$, reads

\begin{equation}\label{general-momentum}
n_\alpha m_\alpha {D{\bf{v}}_\alpha \over Dt}=-\nabla p_\alpha+e_\alpha n_\alpha({\bf{E}}+{\bf{v}}_\alpha\times{\bf{B}})-n_\alpha m_\alpha \nu_{\alpha\beta}(  {\bf{v}}_\alpha-{\bf{v}}_\beta )-n_\alpha m_\alpha \nu_{\alpha\gamma}({\bf{v}}_\alpha-{\bf{v}}_\gamma),
\end{equation}
where $\nu_{\alpha\beta}$ represents the frequency of collisions experienced by the species $\alpha$ from species $\beta$ with $\alpha\neq\beta\neq\gamma$. Here, $n_\alpha$ is the number density and $m_\alpha$ is the mass of the species $\alpha$ and for the other terms we have followed the usual textbook notation. For example, in writing the momentum equation for electrons we may choose $\alpha=e$ with $\beta=i$ and $\gamma=n$. Assuming qusi-neutrality, $n_i=n_e$, we define the electric current as ${\bf{J}}=n_e e ({\bf{v}}_i-{\bf{v}}_e)$ and plasma velocity as ${\bf{v}}=(m_i{\bf{v}}_i+m_e {\bf{v}}_e)/(m_i+m_e)$. For now, let us ignore the neutrals and add the momentum equations of ions and electrons. We obtain the general form of the Ohm's law (Vasyliunas 1975):
\begin{equation}\label{Ohm2}
 {\bf{E}}+{\bf{v\times B}}=\eta {\bf{J}}+{\bf{E}}_H+{1\over n_ee}{\bf{J\times B}}-{1\over n_ee}\nabla p_e+{1\over\omega_{pi}^2}{d{\bf{J}}\over dt},
 \end{equation}
 where $\eta$ is the electrical resistivity originating from electron-ion collisions, $n_e$ is the number density of electrons with the associated electric charge $e$ and pressure $p_e$ and $\omega_{pi}$ is the plasma frequency. Here we have "added" an electric field $E_H$ which originates from hyper-resistivity discussed below. At scales larger than the ion inertial length $\delta_i=c/\omega_{pi}$, with ion plasma frequency $\omega_{pi}$, the last four terms in the RHS become negligible and we recover eq.(\ref{Ohm1}) which describes the resistive MHD limit (Shay et al. 2001). Recall that the ion inertial length is the scale at which the ions in the plasma start to decouple from the electrons and consequently the magnetic field becomes only frozen into the electrons. In a high plasma-$\beta$, on the other hand, the decoupling of the ions and electrons happens at the ion gyroradius $r_i=\beta^{1/2}\delta_i$ (Zweibel \& Yamada 2009). The fourth and fifth terms in the RHS of eq.(\ref{Ohm2}) correspond, respectively, to the electron non-gyrotropic pressure (related to ambipolar diffusion discussed in Sec.(\ref{ssAmbipolar}) and inertia. These effects are associated with breaking the frozen in condition for electrons. However, their effect seem relatively less important in reconnection events (see Shay et al. 2001 and references therein). Since what interests us here, in an astrophysical context, is the scales larger than the ion inertial length (ion skin depth) thus we will neglect these effects.

The first term in the RHS of the above equation is the normal resistive electric field which leads to the Sweet-Parker model. The second term, ${\bf{E}}_H=\eta_H\nabla^2 \bf{J}$ has been suggested by some authors (see e.g., Strauss 1988) to represent the effect of hyper-resistivity (see also Kim \& Diamond 2001). Anomalous electron viscosity, due to micro-scale field line stochasticity and tearing-mode turbulence are among the suggested phenomena that give rise to hyper-resistivity $\eta_H$ (Huang et al. 2013). If so, neglecting all other effects, the Ohm's law takes the form ${\bf{v\times B}}=\eta_H\nabla^2\bf{J}$, so $V_R\sim \eta_H J/B\delta^2$. Combining this result with the Amp\'ere's law and mass conservation, we find the Sweet-Parker speed with hyper-resistivity:
\begin{equation}\label{hyper-resistivity}
V_R\simeq V_A S_H^{-1/4},
\end{equation}
where $S_H=V_A\Delta^3/\eta_H$ is the hyper-resistive Lundquist number. Even if this would give us high reconnection rates, there are theoretical loop-holes in the underlying calculations.  In fact, the calculation of an electric field of the form $\eta_H\nabla^2 \bf{J}$, which we mentioned above, is based on the concept of magnetic helicity and its flux. However, in such calculations some "favorable" terms are kept and manipulated to eventually resemble an enhanced resistivity while the others are simply ignored without a firm theoretical justification.

\subsection{Anomalous Resistivity}\label{ssAmbipolar}

The Spitzer resistivity $\eta$ in a plasma, with negligible ion-neutral collisions, originates from the electron-ion collisions with frequency $\nu_{ie}$:
\begin{equation}\label{19}
 \eta=\nu_{ie} \frac{1}{\omega_{pe}^2},
\end{equation}
where $\omega_{pe}=( ne^2/m)^{1/2}$ is the electron plasma frequency. This is the supposedly constant resistivity employed in the Sweet-Parker model. In fact, the Ohm's law, ${\bf{E+v\times B}}=\eta\bf{J}$, even in its most general form is the momentum equation for the electrons in which the drag force of the collisions is represented by resistivity. Thus one may also look for possible collisions with particles other than ions. It is speculated that small scale electromagnetic fluctuations and waves can also interact with electrons giving rise to some form of "resistivity"---anomalous resistivity. The fluctuations originate form the strong currents and gradients present in the reconnection zone (Zweibel \& Yamada 2016) and lead to micro-instabilities. So in this picture, anomalous resistivity is due to kinetic plasma micro-instabilities and describes the transfer of momentum between electrons and small-scale electromagnetic fluctuations (see also Zweibel \& Yamada 2009; Uzdensky et al. 2009). A relative speed between the ions and electrons larger than the ion acoustic speed in the plasma can lead to a strong anomalous resistivity (Kulsrud 2005). Strong electric currents in the reconnection zone can feed the plasma instabilities which can in turn produce turbulence. 

For a current sheet thickness less than the ion skin depth, the relative motion of ions and electrons in the reconnection zone may increase the reconnection rate by enhancing the resistivity. Suppose that electrons obtain a drift velocity much larger than the average thermal velocity. Then the effective resistivity is much larger than the Spitzer resistivity given in eq.(\ref{19}) (Papadopoulos 1977). This  anomalous resistivity, $\eta_{anom}$ would enhance the reconnection rate in the Sweet-Parker model by a factor of $\sqrt{\eta_{anom}/ \eta}$. This is still too slow (Parker 1979). Assuming that the electron drift velocity, $v_d$, is of order the thermal velocity $u_t=(k_B T/m)^{1/2}$, the condition for the anomalous resistivity to be important is $J>J_{cr}=neu_t$ with the electron number density $n$ and critical current $J_{cr}$. The width of the current sheet, where there is a change of $\tilde{ B}$ in magnetic field, is $\delta \sim  {\tilde{ B}}/{J}$. As $J$ becomes greater than $J_{cr}$, the effective resistivity increases nonlinearly by broadening the current sheet. When $J$ is of order $J_{cr}$, we get
$\delta \sim { \tilde{ B}}/{ n e u_t}$. In terms of the cyclotron radius, $r=mu_t/eB$, where $B$ is the total magnetic field including any shared component, we find

\begin{equation*}
\delta \sim \frac{r \tilde {B}}{B} \left (\frac{V_A}{u_t}\right)^2,
\end{equation*}

which, in combination with mass conservation yields

\begin{equation}\label{22}
V_R \simeq V_A  \; \frac{\tilde {B}}{B} \; \left (\frac{V_A}{u_t}\right)^2 \; \frac{r}{\Delta}.
\end{equation}

If $\Delta \gg r$, this reconnection velocity is much less than  the local Alfv\'en speed. In general, the anomalous resistivity would matter in the reconnection region if the width of the current sheet in the Sweet-Parker model, $\delta_{SP}=(\eta \Delta/V_A)$, is less than $\delta$,

\begin{equation}\label{23}
\delta_{SP} \sim \left (\frac{\eta \Delta}{V_A}\right) < \delta.
\end{equation}

This condition can be easily satisfied for laboratory plasmas but not the interstellar medium if the Sweet-Parker model is used (Lazarian \& Vishniac 1999). The other effect of the anomalous resistivity, apparently, is to keep the current sheet thickness as thick as the ion skin depth. A reconnection layer thicker than that of the Sweet-Parker model leads to a faster outflow of the plasma. Also, this enhanced resistivity may increase the field line breaking rate at the X point and would allow the electrons to flow across the field lines (Yamada et a. 2010). However, in spite of some evidence indicating the existence of anomalous resistivity, yet there is no plausible theory explaining it in terms of an instability. On the other hand, most of the proposed instabilities (Krall \& Liewer 1971; Wang et al. 2008; McBride et al. 1972) are not appropriate for the reconnection problem (for a more detailed discussion see Yamada et al. 2010).

\subsection{Collisionless Reconnection: Hall Effect}

The third term in the RHS of eq.(\ref{Ohm2}) is the Hall term. Ions carry most of the plasma while the electrons are frozen into the magnetic field. So in this case, the ions move in much wider channel enhancing the reconnection rate. This effect, and the associated Whistler dynamics, might be important in collisionless plasmas present in the low density accretion discs and molecular clouds. Also increasing the number of neutrals will push the Hall effect to larger scales and consequently the ion inertial length $\delta_i$ increases by a factor of the square root of neutral to ion ratio, $(n_n/n_i)^{1/2}$ (Zweibel \& Yamada 2009). The work on the effects of the Hall term on reconnection is not yet conclusive (Shay et al. 2001; Wang et al. 2001; Smith et al. 2004). Although the Hall term does not add any unknowns to the MHD equations but it does increase the requirements for spatial and temporal resolution (Yamada et al. 2010). 
\begin{figure}
\includegraphics[scale=.3]{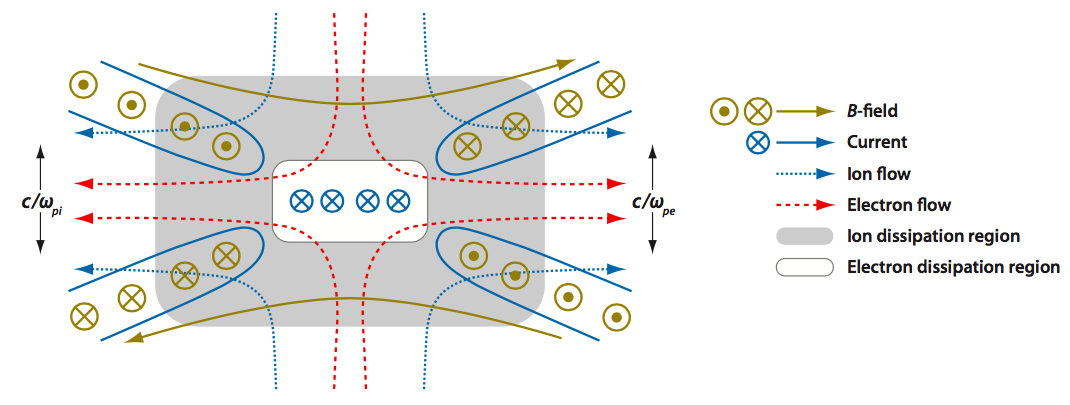}
\centering
\caption {\footnotesize {Collisionless magnetic reconnection. At a distance of order the ion inertial length $\delta_i=c/\omega_{pi}$, ions decouple from the electrons in the current sheet. The electrons flow inward and the magnetic field lines are reconnected in the much thinner electron diffusion layer From Zweibel \& Yamada (2009).}}
\label{collisionless}
\end{figure}

In a collisionless plasma, the equation of motion for a particle of species $\alpha$ of mass $m_\alpha$ and electric charge $e_\alpha$ reads $m_\alpha \partial_t{\bf{v}}_\alpha=e_\alpha({\bf{E}}+{\bf{v}}_\alpha\times{\bf{B}})$. Rearrangement of the terms gives us a more convenient form
\begin{equation}
{d{\bf{v}}_\alpha\over dt}+{\omega}_{\alpha}\times{\bf{v}}_\alpha={e_\alpha\over m_\alpha}\bf{E},
\end{equation}
where $\omega_\alpha=e_\alpha B/m_\alpha$. A comparison of the time scales involved in the above expression, related to the inertial time scale $\tau$ and the gyrofrequency $\omega_\alpha$, shows that if $\tau^{-1}=\omega\ll\omega_\alpha$ the second term in LFH will dominate. In this MHD limit, the magnetic field lines are frozen into the charged particles, both ions and electron, whose motions are an $\bf{E\times B}$ drift. On the other hand, if $\tau^{-1}=\omega\gg\omega_\alpha$, the MHD approximation fails and with the domination of the inertia term, the first term in the above equation, the charged particles start to decouple from the field lines (Zweibel \& Yamada 2009). We can add the equation of motion for electrons of mass $m_e$ and number density $n_e$ to that of ions of mass $m_i$ and number density $n_i$ to get an equation in terms of the electric current. Assume quasi-neutrality $n_e=n_i$, define the current as ${\bf{J}}=en_e({\bf{v}}_i-{\bf{v}}_e)$ and plasma velocity as $m_i{\bf{v}}_i+m_e{\bf{v}}_e/(m_e+m_i)$. We obtain eq.(\ref{Ohm2}) without resistive terms which come from collisions and diffusion processes. This is the collisionless regime. Numerical simulations show that the collisionless reconnection can be faster than the Sweet-Parker reconnection (Zweibel \& Yamada 2009). 

It is usually thought that the Hall effect is dominant in the collisionless regime. When the $\bf{J\times B}$ term dominates, the in-plane flow of electrons into and out of the current sheet generates an electric current parallel to the reconnection zone unlike the collisional Sweet-Parker and Petschek models in which the electric current is perpendicular to the current sheet (Zweibel \& Yamada 2009); see Fig.(\ref{collisionless}). Since the Hall effect is important over the scales less than the ion inertial length, which is proportional to the electron mean free path $c/\omega_{pi}\propto \lambda_{mfp}^{-1/2}$, we can get a condition by comparing this length scale to the width of the Sweet-Parker current sheet $\delta_{SP}\propto L^{1/2}$. We find
\begin{equation}\notag
{\delta_{SP}\over \delta_i}\propto \Big({L\over \lambda_{mfp}}\Big)^{1/2}.
\end{equation}
Therefore, the Hall effect is only important for current sheets whose length $L$ is comparable to the electron mean free path $\lambda_{mfp}$. For this reason, the Hall reconnection is sometimes called collisionless reconnection despite the fact that the latter is a more general mechanism. Some numerical simulations have shown that the collisionless reconnection can be fast even without the Hall term (see e.g., Karimabadi et al. 2004). The problem remains unsettled. Indeed, despite the fact that small scale and hot systems are more likely to follow collisionless reconnection but the details are not well understood. The fact that the most plasmas in astrophysics are turbulent makes it even much difficult to see how, for example, the Hall term may affect the reconnection (for a more detailed discussion of collisionless reconnection see Zweibel \& Yamada 2009 and references therein).

\subsection{X-geomtry: Petschek Model}

\begin{figure}
\includegraphics[scale=.3]{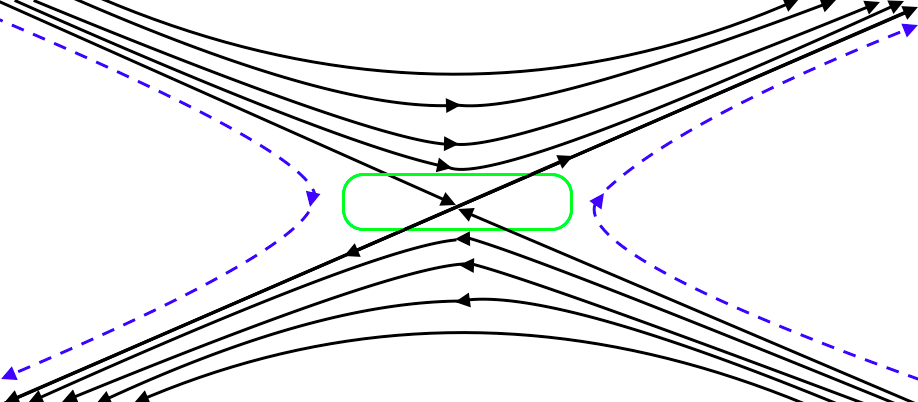}
\centering
\caption {\footnotesize {Petscheck X-point geometry.}}
\label{Petschek}
\end{figure}

As a solution to the problem of slow reconnection rate given by the Sweet-Parker model, Petschek (1964) considered plasma compressibility and invoked a slow shock structure (Guo et al. 2012; Yamada et al. 2010). The formation of slow shocks (see e.g., Ugai \& Tsuda 1977 and Sato \& Hayashi 1979) can widen the Sweet-Parker's narrow channel through which large amounts of plasma need to flow. Consequently, in the so-called Petschek model, the magnetic field lines do not reconnect along the entire length of the current sheet $\Delta$. Instead, reconnection would occur over a shorter length $\Delta^\prime<\Delta$. This is also the distance from the X-point (also called neutral point where total field vanishes) at which the shocks start to propagate; see Fig.(\ref{Petschek}). In this picture, reconnection through a shorter but wider current sheet leads to super-Alfv\'nic outflow speeds which produce shocks. In a highly collisional plasmas, the current sheet resembles a rectangular Sweet-Parker reconnection layer while in collisionless fluids its shape resembles a double wedge similar to the Petschek model (Yamada et al. 2010). In the latter regime, other effects most notably the Hall effect may become important as well. Petschek (1964) showed that the involved MHD equations are independent of this distance so it can be, in principle, very small. In this model, also known as X-point geometry model, the rest of the boundary layer consists of shocks that accelerate the remaining matter which otherwise could not pass through the layer. For highly collisional plasmas, both observations and numerical simulations suggest a rectangular geometry for the current sheet similar to that of the Sweet-Parker model. However, the situation changes dramatically as the collisionality changes. In the collisionless plasmas, an X-point geometry, similar to the Petschek's double wedge, is identified with a much faster reconnection rate (Yamada et al. 2010). 

In the Petschek model, a large reconnection angle forms if the resistivity has its maximum value near the X-point and rapidly decreases away. This in turn speeds up the reconnection. In the Petschek model, unlike the Sweet-Parker scheme, the magnetic energy is mostly converted to the kinetic energy of the ions in the outflow and also to the heat if there are plasma shocks. Thus, in fact, a small amount of the magnetic energy is consumed to resistively heat the electrons (Zweibel \& Yamada 2009). For a current sheet with thickness $\delta $ along the x-axis and the length $\Delta $ along the y-axis, the opening angle for the magnetic field is given by $\tan \theta=B_x/B_y=V_{rec}/V_A$ (Zweibel \& Yamada 2009). Thus, increasing the angle $\theta$ leads to an increase in the reconnection speed $V_{rec}$. As showed before, in the Sweet-Parker model, $\tan \theta \propto S^{-1/2}$ and the reconnection speed is inversely proportional to the current sheet length $V_{rec}\propto \Delta ^{-1/2}$. This means that for a shorter current sheet of length $\Delta' < \Delta$, the reconnection speed enhances by a factor of $(\Delta/\Delta')^{1/2}$. Thus, the Petschek's reconnection speed is given by
\begin{equation*}
V_R=\frac{V_A}{\sqrt{S}} \left ({\frac{\Delta}{\Delta^\prime}} \right)^{1/2}.
\end{equation*}
This result is faster than the Sweet-Parker reconnection velocity by a factor of $\sqrt{\Delta/\Delta^\prime}$. Estimating $\Delta'$, we arrive at an upper limit for the reconnection speed, the so-called Petschek speed, given by

\begin{equation}\label{17}
V_R={\pi\over 8}\frac{V_A}{\ln S}.
\end{equation}
This fast velocity differs from the original result that Petschek formulated by a factor of $2$ which results from the correction of a minor error in Petschek's paper (Yamada et al. 2010; see also Vasyliunas 1975 and Priest \& Forbes 2000). Despite predicting a fast velocity of order the Alfv\'en velocity, this model has been very controversial (Biskamp 2000). This is partly because the origin of the slow shocks was not clear in this model and overall it is not clear how the shocks can form in a self-consistent manner (Yamada et al. 2010). In fact many numerical simulations confirmed that the involved geometry is not stable. It also turns out that $\Delta^\prime$ is not a free parameter, in contrast to what Petschek showed, and there is an additional condition that Petschek did not include in his model (Kulsrud 2000). Numerical simulations performed by Biskamp (1986) confirmed the generation of shocks, however, only at a distance comparable to the global scale of the system---much larger than the value predicted by Petschek's model (Yamada et al. 2010).

In fact, these simulations and those performed subsequently by Uzdenski and Kulsrud (2000) demonstrated the validity of the Sweet-Parker scheme for constant resistivity (see also Guo et al. 2012). Numerical simulations, however, have confirmed Petschek-like reconnection for spatially variable resistivities (Malyshkin et al., 2005). In fact, Petschek reconnection would not initiate and proceed by itself and, if even imposed initially, would be unstable unless the resistivity increases near the neutral X-point (Zweibel \& Yamada 2009). In fact, in numerical simulations with constant resistivity, one can impose an X-point geometry as an initial condition to enforce Petschek-type reconnection (Zweibel \& Yamada 2009), however, it has been shown that it would revert to the Sweet-Parker scheme (Uzdensky \& Kulsrud 2000). On the other hand, if resistivity is spatially variable and increases in the current sheet, the Petschek's reconnection can be stable (see e.g., Sato \& Hayashi 1979; Scholer 1989; Baty, Priest \& Forbes 2006; Zweibel \& Yamada 2009). Granted, only a spatially variable resistivity can make the otherwise unstable Petschek reconnection survive, but the Petschek model provides the most powerful and the fastest scheme among the conventional models. On the other hand, its unstable geometry and incompatibility with the numerical simulations have gradually led to consideration of other models. 

In short, numerical simulations performed to produce Petscheck-like structures are not promising (Biskamp 1984, 1986). Typically what happens is that the X-point region rapidly collapses toward the Sweet-Parker geometry with increasing the Lundquist number (see also Wang, Ma, \& Bhattacharjee 1996). We can consider perturbations of the original X-point geometry to understand the nature of this collapse. To keep the original geometry stable, reconnection should be fast which would require shocks in the original model. On the other hand, the shocks, supported by the flows driven by fast reconnection, would fade away $L_x$ increases. One would expect that generally, and at least in the collisional regime, reconnection occurs through narrow current sheets (LV99). There have been other approaches as well. For example, Vishniac (1995a, 1995b) resorted to a different approach in high-$\beta$ plasmas based on the notion of narrow flux tubes with concentrated magnetic field. This allows one to reduce the length scale $L_x$ to the flux tube radius. This way, one can obtain fast reconnection rates even with the Sweet-Parker geometry. (For a more detailed review of the Petschek model see Yamada et al. 2010).

\subsection{Ambipolar Diffusion}

In a plasma with neutrals, we have to take care of the neutral-ion and probably neutral-electron collisions as well. The equation of motion for each species is given by eq.(\ref{general-momentum}) with the same definition of the electric current and plasma velocity. In the steady state, we can ignore all other forces but the Lorentz and collisional forces, and combine all three equations for neutrals, electrons and ions to get
\begin{equation}\label{drift}
{\bf{J\times B}}=n_i m_i \nu_{in}({\bf{v}}-{\bf{v}}_n)\simeq \rho_i \nu_{in}({\bf{v}}_i-{\bf{v}}_n),
\end{equation}
where $\rho_i=n_i m_i$ is the ion mass density and $\nu_{in}$ is the ion-neutral collision frequency. We have assumed that ions are much heavier than electrons so their velocity is very close to the plasma velocity, $v_i\simeq v$. The ion-neutral drift velocity ${\bf{v}}_D={\bf{v}}_i-{\bf{v}}_n $ leads to the concept of ambipolar diffusion in which the field moves along with the ions in the background of the neutrals. We can now write the induction equation, $\partial_t{\bf{B}}=\nabla\times({\bf{v\times B}}-\eta \bf{J})$, that after using eq.(\ref{drift}) becomes
\begin{equation}\label{ambipolar}
{\partial {\bf{B}}\over \partial t}=\nabla\times \Big({\bf{v}}_n\times{\bf{B}}+{   {\bf{J.B}}   \over  \rho_i\nu_{in}  }-(\eta+\eta_{am}) {\bf{J}}\Big),
\end{equation}
where $\eta_{am}=B^2/\rho_i\nu_{in}$ is the ambipolar diffusion coefficient which is a variable magnetic diffusivity. Also we note that the drift velocity ${\bf{v}}_D$ has a component proportional to the magnetic pressure gradient, $-\nabla B^2$. Thus, magnetic flux will be pushed toward the magnetic null points. At these points, on the other hand, $\eta_{am}$ vanishes which prevents magnetic field from smoothing out and makes the null points singular (Brandenburg \& Zweibel 1994). 

\begin{figure}
\includegraphics[scale=.4]{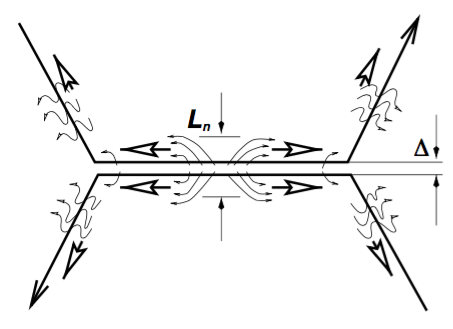}
\centering
\caption {\footnotesize {A reconnection layer of width $\Delta$. The vertical velocity of matter inflow is the reconnection speed whereas the horizontal outflow velocity is typically of order the local Alfv\'en speed (see the text). Ion velocities, marked with thick lines, and neutral velocities, marked with thin lines, are appropriate for ambipolar diffusion. The outflow of gas is confined to the region of width $L_n$ (illustration from Vishniac \& Lazarian 1999).}}
\label{VL-ambipolar}
\end{figure}

%\iffalse

Vishniac and Lazarian (1999) estimated the reconnection speed in different phases of the interstellar medium (ISM), using the Sweet-Parker geometry, emphasizing the role of neutral particles; see Fig.(\ref{VL-ambipolar}). The main effect of the neutral drag is thought to be driving the transverse ejection of the neutrals without any significant reduction in ion velocities. The authors argued that, in the ISM, the recombination rate is critical in limiting the reconnection speed. They found the ambipolar diffusion provided reconnection speeds much faster than the conventional Sweet-Parker speed but still too slow compared with the local Alfv\'en speed; 
\begin{equation}
V_{rec}\sim {1\over x\beta}\Big( {\eta\over\tau}\Big)^{1/2} (1+2 x \beta)^{1/2},
\end{equation}
where $x$ is the ionization fraction in the ISM, $\tau$ is the recombination time scale and $\beta$ is the ratio of the gas pressure to magnetic pressure. This yields a rate higher than the Sweet-Parker rate in the colder parts of the ISM (typically with temperatures lower than $T\sim 100$K). However, at most, in the dense molecular clouds, the speed hardly exceeds $10^{-3} V_A$. Their corrected Sweet-Parker speed, in an ambipolar medium with a long recombination time, is given by
\begin{equation}\label{VL1}
V_{rec}\simeq \Big({2\eta_\infty\over \tau_{recomb,\infty}}\Big)^{1/2}\Big({T_\infty\over T(0)}\Big)^{1/2}\Big[\Big( {T_\infty\over T(0)}\Big)^2\Big(1+{1 \over \beta x} \Big)^2-1 \Big]^{1/2}.
\end{equation}
where $\eta_\infty$ and $T_\infty$ are the resistivity and temperature at infinity and $T(0)$ is the temperature at the reconnection layer. Also note that the speed given by eq.(\ref{VL1}) is independent of $L$. However, eq.(\ref{VL1}) is obtained only for the case when the recombination time, $\tau_{recomb}$, is much less than the shearing time $L/V_A$. Otherwise, in the opposite limit, ions escape from the sides of the reconnection region rather than through recombination. The corrected Sweet-Parker speed in this case is given by
\begin{equation}\label{VL-corrected}
V_{CSP}\simeq \Big( {V_A\eta_\infty\over L} \Big)^{1/2}\Big( {\eta_\infty\over \nu_n} \Big)^{1/8}\Big({T_\infty\over T(0)} \Big)^{21/16}\Big( 1+{1\over x\beta}\Big)^{3/8},
\end{equation}
 where $\nu_n$ is the viscosity of the neutral gas. Note that this speed depends on $L$ and reconnection occurs with higher speeds on smaller distances. 
 
Ambipolar diffusion has been studied for decades (see e.g., Mestel 1985). It has been suggested that the ion-neutral drift would be faster in astrophysical turbulent media (Zweibel 2002). Numerical work (Zweibel \& Brandenburg 1995; Brandenburg \& Zweibel 1997) has shown that ambipolar diffusion may lead to very narrow reconnection regions whose widths are determined by the size of particle orbits in the plasma (Viashniac \& Lazarian 1999). However, overall, it seems unlikely that the ambipolar diffusion leads to fast reconnection rates observed in astrophysics.

\iffalse

\begin{figure}
\includegraphics[scale=.6]{VL-ambipolar2.png}
\centering
\caption {\footnotesize {Reconnection speeds obtained by Vishniac and Lazarian (1999) for the ionized phases of the ISM given in cgs units. $n$ is the number density and $\tau_{recomb}$ is the recombination time outside the reconnection zone; $V_{max}$ follows from infinitely fast recombination; $V_{rec}$ allows for a realistic recombination rate; $V_{SP}$ is the Sweet-Parker reconnection speed; and $V_{CSP}$ is the corrected Sweet-Parker reconnection speed given by eq.(\ref{VL-corrected}). $V_{rec}$ for $H_2$ is placed in parentheses since it allows for heating of the reconnection layer. The Alfv\'en speed is of order $10^5$ cm/s.}}
\label{VL-ambipolar(2)}
\end{figure}

\fi

\section{Reconnection in Turbulent Media} \label{s3}

\begin{figure}
 \begin{centering}
\includegraphics[scale=.27]{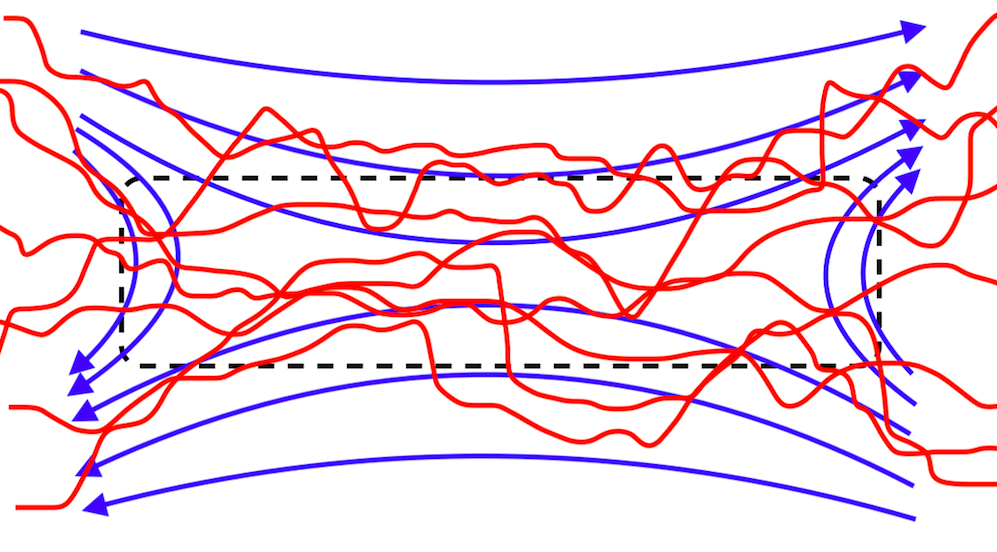}
\includegraphics[scale=.45]{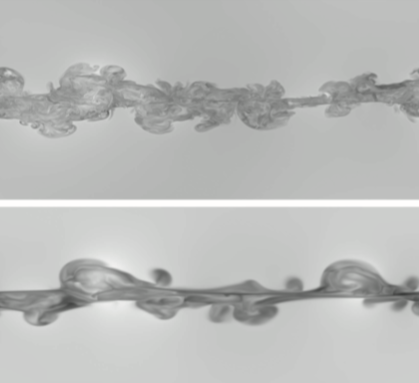}
\caption {\footnotesize {Left: Fiducial large scale (blue curves) and small scale (red curves) magnetic field lines in a turbulent reconnection zone. Turbulence can be present initially in the medium or produced later on by current sheet instabilities. The field lines show a fuzzy topology that suggests a diffusional approach. The ejection velocity is expected to be of order the Alfv\'en speed $V_{eject}\sim V_A$ unless a large viscosity slows down the outflow speed; see the text. Right: Turbulent current layer with the mean magnetic field out of the plane, from numerical simulations performed by Beresnyak (2017). The upper panel is the magnitude of large scale magnetic field in the hyper-viscous simulation, and the lower panel is the same in the viscous simulation.}}
\end{centering}
\label{Stochastic-sheet}
\end{figure}

Astrophysical systems are generally turbulent. Applying conventional reconnection models in such environments overlooks the effects of turbulence either by completely ignoring it, e.g., employing the original Sweet-Parker model, or else failing in taking its major effects into account, e.g., relying on indirect effects of turbulence like enhanced anomalous diffusivity. Even in the absence of any external turbulence, current sheet instabilities can generate self-sustained internal turbulence that can affect reconnection as indicated by 3D numerical simulations (Beresnyak 2013; 2016; Oishi et al. 2015; Kowal et al. 2016; Huang \& Bhattacharjee 2016). There is, in fact, a subtler point here about the current sheet instabilities; see Fig.(\ref{Stochastic-sheet}). One forgotten question, to be answered or at least to be asked, in the Sweet-Parker and Petschek models is that how, and under what conditions, reconnection initiates and occurs: Why should two plasma regions with oppositely oriented magnetic fields initiate reconnection process? Furth et al. (1963) showed that this configuration is unstable to small perturbation and leads to tearing instability which eventually reconnects the field lines (Zweibel \& Yamada 2009). At the same time, the instability will generate turbulence which itself affects an ongoing reconnection process. Formation of plasmoid structures in thin reconnection regions have been observed in resistive MHD simulations in the high Lundquist numbers (Samtaney et al. 2009; Huang \& Bhattacharjee 2010). It is thought that this may lead to fast reconnection (Guo et al. 2012; Shibata \& Tanuma 2001; Drake et al. 2003, 2006). Here, we focus on tearing modes just as an example to see how instabilities can affect reconnection (see also Strauss 1988). We also go over some main results of the GS95 MHD turbulence which will be useful for us in the later sections. 

The original Sweet-Parker and Petschek models dealt only with the steady state magnetic reconnection. In astrophysical plasmas, as long as very long time scales are considered, the ideal MHD approximation breaks down and the magnetic fields diffuse into the plasma. Although resistivity seems to damp out any perturbation but in many cases it destabilizes the plasma. It can be shown that the magnetic field, in general, is unstable to small perturbations (Yamada et al. 2010). The key point in understanding how resistivity can have a destabilizing role is that it breaks the frozen-in constraint and allows plasma to separate from the magnetic field which gives rise to a new class of instabilities. Current sheets are unstable to tearing modes (Furth et al. 1963) and the absence of external turbulence would not necessarily lead to the Sweet-Parker reconnection (Jafari et al. 2018). Non-zero resistivity, electron inertia or electron shear viscosity can all decouple the magnetic field form the plasma which in turn leads to the tearing instability. The non-uniformities of magnetic field is the driving force of the tearing instability which arises when the resistivity is non-zero but can persist when it tends to zero. The situation is similar to the case of the Taylor-Rayleigh instability in which the driving force is the non-uniform plasma pressure.\\ 

\subsection{Turbulent Diffusion}

Turbulence is generated by the internal instabilities in the reconnection zone and can be present in the environment before reconnection occurs. Reconnection itself generates turbulence. No matter how it is generated, turbulence leads to an "enhanced" magnetic diffusivity. With eddy scale $l$ and velocity $\delta v$ in the turbulent cascade in a large scale reconnection zone, one might naively think of turbulent effects just in terms of an enhanced diffusivity $\eta_t\equiv \beta \sim l \delta v\sim \delta v^2/\tau$ where $\tau$ is the eddy turn-over time. The back reaction through the Lorentz force affects the magnetic dynamos in a process known as $\alpha$-quenching. In a turbulent medium, a possible dynamo action relies on the generation of a turbulent electric field $\cal{E}$ which can be expanded in terms of the large scale magnetic field $\bf{B}$ and electric current $\bf{J}$ as ${\cal{E}}=\alpha {\bf{B}}-\beta \bf{J}$ plus probably other terms related to rotation and shear. The quenching of $\alpha$ destroys the dynamo action and is the subject of intense research in dynamo theories (see e.g., Moffatt 1978; Krause \& R{\"a}dler 1980). A similar quenching phenomenon has also been suggested for the diffusivity $\beta$ (Kim \& Diamond 2001). 

The replacement of normal (Spitzer) resistivity with an enhanced turbulent resistivity, however, lacks a firm theoretical justification. In fact, such a simple treatment of a complex phenomenon like turbulence seems extremely unrealistic. Matthaeus \& Lamkin (1985, 1986), through 2D numerical simulations, found no evidence that turbulence could affect the average reconnection rate. However, tearing-modes induced turbulence was proposed to affect the reconnection through hyper-resistivity, discussed earlier. This might increase the reconnection speed over scales much larger than the turbulent scale (Strauss 1988). On the other hand, Kim \& Diamond (2001) argued that the back reaction of the magnetic field on the turbulent cascade could suppress turbulent diffusion (see also Zweibel \& Yamada 2009).  But at this point, we need to remind ourselves of the important fact that, in a turbulent environment, the magnetic field lines show a "stochastic" behavior and the nature of their diffusion changes. In turbulent media, frequently encountered in astrophysics, the separation of neighboring fluid elements grows very rapidly with time. In such a noisy environment, the notion of "following a fluid element" has no meaning, and consequently, the notion of a magnetic field frozen to the fluid elements loses its meaning as Zweibel \& Yamada (2016) have pointed out. This is spontaneous stochasticity (LV99; Eyink et al. 2013). Also, it turns out, that there is a huge difference between two and three-dimensional turbulence. Instead of a normal diffusion in which the r.m.s separation of field lines grows linearly with time, in the inertial range of 3D turbulence, it becomes a cubic function of time (Richardson diffusion) as discussed before. Therefore, a simple replacement of "diffusion coefficient" in a typical diffusion process would not work. Only in the last two decades, more realistic turbulent models have been proposed (for a detailed review see e.g., Lazarian et al. 2015). In \S 5, we will study a completely different approach, called the stochastic reconnection (see also Lazarian \& Vishniac 1999; Eyink 2011; Eyink et al. 2011).

\subsection{Current Sheet Instabilities}

When two regions in a plasma with opposite magnetic fluxes come into contact, some magnetic islands will form. In the presence of a shared field component between the two regions, instead of magnetic islands, non-linear Alfv\'enic waves will form which differ from the islands although their projection to the plane of oppositely oriented fields would look very similar to the islands. However, while in the former case the instability starts with a linear growth and evolves into a non-linear regime, in the latter case, the instability stagnates as it enters its non-linear stage (Lazarian \& Vishniac 1998). The tearing modes instability reduces the magnetic energy by partly converting it into the other forms, e.g., the thermal energy of electrons. Relaxing the magnetic field configuration to another configuration with lower energy causes magnetic reconnection. The growth rate of the tearing modes at lower wavenumbers is given by (Furth et al. 1963)
\begin{equation}\label{TM-rate}
\gamma=\Big({S\over ka}\Big)^{2/5}{\eta\over a^2},
\end{equation}

 \begin{figure}
 \begin{centering}
\includegraphics[scale=.5]{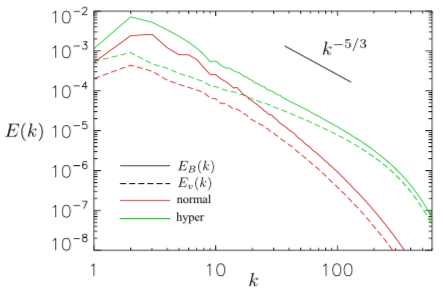}
\includegraphics[scale=.5]{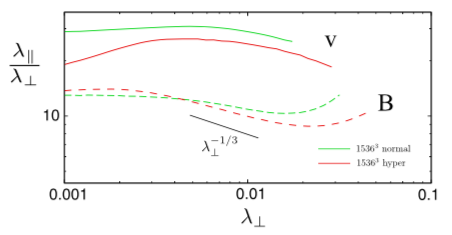}
\caption {\footnotesize {Left: Velocity and magnetic power spectra. The spectral slopes were around $-1.5$  and $-1.7$, which is characteristic of local-in-scale turbulence. Right: Anisotropy of velocity and magnetic perturbations in the current layer measured with respect to the local field. Both panels from Beresnyak (2017).}}
\end{centering}
\label{Beresnyak2-3}
\end{figure}

where $S=V_A a/\eta$ is the Lundquist number, $k$ is the transverse wavenumber and $a$ is the thickness of the current sheet. The maximum growth rate is achieved for the minimum wavenumber of the growing modes. The upper and lower bounds of the wavenumber is given by $S^{-1/4}<ka<1$.  This leads to the fastest rate $\gamma\sim S^{1/2}{\eta/ a^2}$ which also corresponds the fastest reconnection speed. This growth rate for the tearing modes depends on the geometric mean of the Alfv\'en and diffusive times and is still slow if we compare it with the observations of, for example, solar flares (Browning \& Lazarian 2013). However, there has been some controversy on the minimum wavenumber. Van Hoven and Cross (1971) argued that the above solution assumes infinite magnetic fields far from the origin (LV tearing) and instead suggested an upper limit given by $ka>S^{-3/7}$ which corresponds to the growth rate (Lazarian \& Vishniac 1998)
 \begin{equation}\label{LV10}
 \gamma \simeq S^{4/7} \eta/a^2.
 \end{equation} 

Any component of the magnetic field shared between the two regions will cause this unstable layer to eject the plasma with a velocity of order the local Alfv\'en speed $V_A$. Consequently, we get a local transverse shear of order $V_A/L$ with $L$ being the length of the layer. If this shear exceeds the tearing instability rate $\gamma$, it will suppress the instability which otherwise mixes the differently oriented magnetic fields. Using this, Bulanov, Salaki and Syrovatskii (1979; see also Lazarian \& Vishniac 1998) suggested the equilibrium condition as
\begin{equation}
{V_A\over L}\sim \gamma,
\end{equation}
where $\gamma$ is meant to be the maximum growth rate of the instability. Assuming the reconnection rate to be $a\gamma$, with $a$ as a characteristic scale of the amplitude of the most rapidly growing tearing perturbations, Lazarian and Vishniac (1998) argued that eq.(\ref{LV10}) would yield ${a/ L}\simeq S^{3/7}$ that leads to the reconnection speed
\begin{equation}\label{LV11}
V_{rec}:= a\gamma\simeq V_A R_e^{-3/10},
\end{equation}
where the magnetic Reynolds number is defined as  $R_e={L V_A/ \eta}={LS/ a}$. Eq.(\ref{LV11}) is the main result achieved by Lazarian and Vishniac (1998) which is faster than the Sweet-Parker rate $\sim R_e^{-1/2}$. They found that the reconnection, with tearing instability, persists unlike the conventional Sweet-Parker model even when the resistivity tends to zero. In the case of a vanishing resistivity, $\eta$ should be replaced with the electron inertia term $\omega(c/\omega_p)^2$ where $\omega_p$ is the plasma frequency and $\omega\sim \gamma$. Using $\gamma\simeq V_A/L$, this yields a reconnection speed of order 
\begin{equation}
V_{rec}\simeq V_A\Big({c\over L\omega_p}\Big)^{3/5}.
\end{equation}
This is the minimal reconnection speed when $\eta$ tends to zero. Strauss (1988) concluded that tearing instability would lead to a fast reconnection of order the Alfv\'en speed. However, this is controversial since although the tearing modes may grow to the point of marginal non-linearity, however, as Lazarian and Vishniac (1998) pointed out, this does not necessarily leads to fast reconnection as other global constraints affect the reconnection speed in different ways. The conclusion is that the tearing instability in an inviscid plasma is not a promising way to enhance the reconnection speed, however, the instability may produce internal turbulence. 

Including viscosity leads to a more complicated situation. Since magnetic reconnection is basically a three dimensional phenomenon, we need to consider the effects of a guid field---third component of the magnetic field---which may affect the reconnection in a major way (Kivelson \& Russell 1995; Ono et al. 1993; Brown 1999; Cothran et al. 2003). A shared field component might also reduce the viscosity in the current sheet (Glukhov 1996). So instabilities in the reconnection region are probably best understood using three dimensional treatments. In the presence of viscosity, the scaling $\gamma\propto \eta^{3/5}$ should be replaced by $\gamma\propto \eta^{2/3}Pr_m^{-1/6}$ (Porcelli 1987). Numerical simulations have confirmed this scaling (see e.g., Takeda et al. 2008).

Current sheet instabilities have been explored in several recent numerical simulations (see e.g., Beresnyak 2013; 2016; Oishi et al. 2015; Kowal et al. 2016; Huang \& Bhattacharjee 2016). All these simulations predict reconnection speeds of the order of $0.01 V_A$ with modest, or negligible, dependence on the resistivity. Nevertheless, some of the numerical results differ in significant ways (Jafari et al. 2018). The simulations performed by Beresnyak (2013; 2016), Oishi et al. (2015) and also Kowal et al. (2016) all show the growth of anisotropic turbulence in the reconnection channel with a power spectrum consistent with the inertial cascade of forced turbulence $E(k)\propto k^{-5/3}$. Huang and Bhattacharjee (2016) found a somehow slower reconnection rate but a much steeper power spectrum, $E_B(k)\propto k^{-2.2}$, suggesting the absence of strong local turbulent cascade. However, their power spectrum is not so steep as to rule out the diffusion of the field lines in the reconnection zone. In all cases the magnetic Prandtl number was unity. All these simulations used closed or periodic boxes so they did not achieve a stationary state with balanced outflows and inflows. But rather, the reconnection rates in these simulations were estimated from the linear growth of the reconnection channel. The relationship between the tearing modes, with resistivity dependent properties, and the resistivity independent reconnection rate remains unclear. Nevertheless, we expect that the current sheet instabilities are able to produce a reconnection speed at least of order $10^{-2} V_A$ and also a turbulent velocity of more or less the same order (Jafari et al. 2018). As we will see in \S 5, this leads to a minimum velocity for the large scale turbulent eddies and also a minimum Reynolds number for eq.(\ref{limit}).

\subsection{MHD Turbulence}

In the Kolmogorov's hydrodynamic turbulence (Kolmogorov 1941), the statistics of turbulent motions is determined by the energy transfer rate $\epsilon=v_y^3/y$ and length scale $y$ at the inertial range, where the energy is supposedly transferred with no dissipation, and the energy transfer rate and viscosity $\nu$ at the dissipative range, where the energy is dissipated by viscosity. Dimensional analysis leads to the turbulent velocity $v_y\sim \epsilon^{1/3}y^{1/3}$ and the turbulent eddy time scale $\tau_y\sim \epsilon^{-1/3}y^{2/3}$ in the inertial range. 
This scaling leads to the famous velocity spectrum $v_k\sim k^{-1/3}$ and energy power spectrum 

\begin{equation}\label{Kscaling}
E_{Kol}(k)\sim k^{-5/3}. 
\end{equation}
At the dissipative range, a similar dimensional analysis shows that $v_d\sim \epsilon^{1/4}\nu^{1/4}$, $\tau_d\sim \epsilon^{-1/2}\nu^{1/2}$ and $y_d\sim \nu^{3/4}\epsilon^{-1/4}\equiv y(\nu/y v_y)^{3/4}$. 

Kolmogorov scaling, for hydrodynamic turbulence, cannot be applied directly to MHD turbulence because of the complications that magnetic field introduces. In the following, we briefly discuss the main ideas in several models of MHD turbulence.\\

\subsubsection{Iroshinkov-Kraichnan Model}

Iroshinkov (1963), and Kraichnan (1965) independently developed a model for the incompressible magnetohydrodynamic (MHD) turbulence on the base of interactions of triads of waves (Diamond \& Craddock 1990). The Iroshinkov-Kraichnan (IK) approach can be summarized as follows. If we assume that in the inertial range with scale $l$ much smaller than the injection scale $L$, and much larger than the dissipation scale $l_{d}$, the interactions are local, i.e. only fluctuations of comparable scales interact, then we can think of propagating fluctuations as Alfv\'en wave packets of parallel extent $l_{\parallel}$ and perpendicular extent $l_{\perp}$. Now, assume $\delta u_{l} \sim \delta B_{l}$ where $\delta u_{l}$ and $\delta B_{l}$ are, respectively, the fluctuations in the velocity and magnetic field. Two counterpropagating wave packets would require an Alfv\'en time, $\tau_{A}$, to pass through each other. During this time, the amplitude of each wave packet will suffer a change $\Delta \delta u_{l}$,
\begin{equation}\label{61-3}
\frac{\Delta \delta u_{l}}{\tau_{A}} \sim \frac{\delta u_{l}}{\tau_{l}},
\end{equation}
where $\tau_{l} \sim l/ \delta u_{l}$ is the eddy time, and $\tau_{A} \sim l_{\parallel}/V_A$ is the Alfv\'en time. The first assumption in the IK theory is weak interactions;
\begin{equation}\label{61-2}
\Delta \delta u_{l} \ll \delta u_{l} \Leftrightarrow \tau_{A} \ll \tau_{l}.
\end{equation}
The cascade time, $\tau_{nl}$ is defined as the time that it takes to change $\delta u_{l}$ by an amount comparable to itself assuming that the changes accumulate in a random walk manner;
\begin{equation}\label{61-1}
\sum \limits_{t=0}^{\tau_{nl}} \; \Delta \delta u_{l} \sim \delta u_{l} \frac{\tau_{A}}{\tau_{l}} \left({\frac{\tau_{nl}}{\tau_{A}}}\right)^{1/2} \sim \delta u_{l}.
\end{equation}
Thus, we obtain the energy transfer rate
\begin{equation}\label{61-0}
\tau_{nl} \sim \frac{l^2 V_{A}}{l_{\parallel} \delta u_{l}^2}.
\end{equation}
The second assumption in Iroshinkov-Kraichnan model is that power should spread isotropically in wavevector space. Assuming $l_{\parallel} \sim l$, we obtain
  \begin{equation}\label{61}
\tau_{IK} \simeq \frac{l V_A}{v_k^2},
\end{equation}
where we used $v_{k}$ instead of $\delta u_{l}$ for simplicity. When $v_k < V_A$, we have $\tau_{nl} > k v_k$, which implies the reduction of the energy cascade to higher wavenumbers by the magnetic field. Constancy of the energy transfer rate $v_k^2/\tau_{IK}=const.$ leads to $v_k\sim k^{-1/4}$. Therefore the Kraichnan-Iroshinkov energy power spectrum is given by 

\begin{equation}\label{IKScaling}
E_{IK}(k)\sim k^{-3/2}.
\end{equation}

\subsubsection{Goldreich-Sridhar Model}

The IK theory was the most popular model accepted as MHD generalization of Kolmogorov ideas for about 30 years. In 1970's, measurements showed strong anisotropies in the solar wind with $l_{\parallel} > l_{\perp}$. Goldreich and Sridhar (1995 henceforth GS95; 1997) suggested that the effect of residual three wave couplings is consistent with eq.(\ref{61-0}), for the basic nonlinear timescale, but an anisotropic spectrum should be considered in which the transfer of  power between modes moves energy toward larger $k_{\bot}$ with no effect on $k_{\parallel}$, by this convention that, $k_{\bot}$ is the wavevector component perpendicular, and $k_{\parallel}$ is the wavevector component parallel, to the direction of magnetic field. Therefore, using eq.(\ref{61-0}), the basic nonlinear timescale can be written as
\begin{equation}\label{62}
\tau_{nl} \simeq \frac{k_{\parallel}V_A}{k_{\bot}^2 v_k^2},
\end{equation}
where $k_{\perp}=l^{-1}$, and $\omega_A=k_{\parallel}V_A$ is Alfv\'en wave frequency.

 The critical balance requires that typical eddies with wavenumbers parallel to the local magnetic field, $k_\|$, and a perpendicular wavenumber, $k_\perp$, are related by the following condition:
\begin{equation}\label{critical-balance}
k_\| V_A\approx k_\perp v_k,
\end{equation}
where $V_A$ is the Alfven speed and $v_k$ is the rms velocity on the eddy scale. This is translated into the requirement that the field couples to a typical eddy at a rate approximately equal to the eddy turnover rate. The second assumption in the GS95 model is that the nonlinear energy transfer rate is $\sim k_\perp v_k$, similar to that of hydrodynamic turbulence (Kolmogorov 1941). These assumptions together lead to a power spectrum which behaves like hydrodynamic turbulence, i.e. $v_k\propto k_\perp^{-1/3}$. Consequently, the energy power spectrum of GS95 is given by

\begin{equation}\label{GS95Scaling}
E_{GS}(k_\perp)\sim k_\perp^{-5/3}.
\end{equation}

Suppose that we inject energy into the medium with a parallel length scale $l$, with the corresponding perpendicular scale $l_{\bot}=l_\parallel M_A$ (where $M_A=V_T/V_A$ is the Alfv\'en Mach number), that creates an rms velocity $V_T$. The resulting inertial turbulent cascade satisfies critical balance at all smaller scales, therefore
\begin{equation}
\tau_{nl}^{-1}\simeq k_\| V_A\simeq k_{\perp}v_k,
\end{equation}
and the constant flow of energy through the sub-Alfv\'enic cascade is given by
\begin{equation}\label{Vrate}
\epsilon\simeq {v_l^4\over l_\parallel V_A}\simeq {V_T^2 V_A\over l_\parallel }\simeq {v_k^2\over \tau_{nl}}\simeq k_\perp v_k^3,
\end{equation}
where $v_l=\sqrt{V_T V_A}$ is sometimes defined as the velocity for isotropic injection of energy which undergoes a weakly turbulent cascade and ends up with a strongly turbulent cascade (see e.g., LV99; Lazarian et al. 2015). Putting all this together we obtain
\begin{equation}\label{67}
k_{\parallel}=\lambda_\parallel^{-1} \simeq l_\parallel ^{-1}\left({k_\perp l V_T\over V_A}\right)^{2/3},
\end{equation}
which is eq.(\ref{67}), and
\begin{equation}\label{68}
\tau_{nl}^{-1} \simeq k_\| V_A\simeq  {V_A\over l_\parallel }\left({k_\perp l V_T\over V_A}\right)^{2/3}.
\end{equation}

The rms velocity in the large scale eddies is 
\begin{equation}\label{69}
 v_k \simeq  V_T\left({k_\perp l V_T\over V_A}\right)^{-1/3}.
\end{equation}

The parallel wavelength at the viscous damping scale, $\lambda_{\parallel,d}=k_{\parallel, d}^{-1}$ can be obtained from eq.(\ref{67}) by substituting $k_{\perp ,d}^{-1}\sim l_\perp(\nu/l_\perp V_T)^{3/4}$. The latter vertical damping scale comes from the Kolmogorov scaling for the viscose damping length scale, $l_d\sim \epsilon^{-1/4}\nu^{3/4}$ with $\epsilon\simeq V_T^3/l_\perp\simeq V_T^2V_A/l_\parallel$ where we have also used $l_\perp V_A=l_\parallel V_T$. If we define the injection velocity as $v_l=\sqrt{V_TV_A}$, then we can write $\epsilon\simeq V_T^2V_A/l_\parallel=v_l^4/(l_\parallel V_A)$ as used in (\ref{Vrate}). 

In passing, we note that one can make a connection between the Kolmogorov scalings, for hydrodynamic turbulence, and MHD turbulence by defining the parallel wavenumber as $k_\parallel^{-1}=\lambda_\parallel \rightarrow V_A \tau_{y}$ and  the perpendicular wavenumber as $k_\perp^{-1}=\lambda_\perp\rightarrow y$. Below the viscous dissipation scale, these become $k_{\parallel, d}^{-1}=\lambda_{\parallel, d} \rightarrow V_A\tau_{ d}$ and $k_{\perp, d}\rightarrow y_d$. Note that the counterpart of the Kolmogorov length scale is the perpendicular wavenumber $y\rightarrow k_\perp^{-1}$ in MHD turbulence. For example, $y_d\sim \nu^{3/4}\epsilon^{-1/4}\rightarrow k_{\perp, d}^{-1}$ and $\epsilon\simeq v_y^3/y\rightarrow V_T^3/l_\perp\simeq V_T^2V_A/l_\parallel$, where we assumed $l_\perp V_A \simeq l_\parallel V_T$ with $V_T$ used in place of $v_y$ as the turbulent eddy velocity. Using the Kolmogorov scalings, and these definitions, we can obtain the parallel wavenumber in GS95 model, given by (\ref{67}).

\subsubsection{Post-GS95 Models}

The MHD model of GS95 has been the subject of numerous numerical and analytical investigations, many of which have been in favor of its $-5/3$ spectral slope (see Beresnyak 2012 and references therein). However, there have been criticisms as well. The second assumption in GS95, for example, has been criticized by Boldyrev (2005; 2006). The elongation of turbulent fluctuations along the large-scale magnetic-field, proposed by GS95 model, has been observed in several numerical simulations of MHD turbulence in a strong external magnetic field (see e.g., Cho \& Vishniac 2001; Milano et al. 2001; M{\"u}ller et al. 2003). Nevertheless, several other simulations have found a different energy spectrum $E(k_\perp)\sim k_\perp^{-3/2}$ (see e.g. M{\"u}ller et al. 2003\&2005; Haugen et al. 2004). Such a spectrum along with the anisotropy of fluctuations is in contradiction with both the Iroshnikov-Kraichnan and the Goldreich-Sridhar MHD turbulence models (Boldyrev 2006).

Boldyrev (2005; 2006) proposed a phenomenological theory of strong incompressible MHD turbulence in the presence of a strong large scale external magnetic field. According to this picture, in the inertial range of scales, fluctuations in magnetic field and velocity field tend to align their polarization directions without reaching a perfect alignment as a result of the presence of a constant energy flux over the scales. In this model, the Alfv\'en waves in the turbulent cascade spontaneously evolve towards parallel polarizations. Consequently, the nonlinear energy transfer rate should include a scale dependent correction term proportional to $k_\perp^{1/4}$. The weakly nonlinear interaction rate is 
\begin{equation}\label{62}
\tau_{nl}^{-1} \simeq {k_{\bot}^2 v_k^2 \over k_{\parallel}V_A }f_{align}(k_\perp),
\end{equation}
where $f_{align}$ is an efficiency factor smaller than unity in Boldyrev's model. So, with $k_{\bot} \gg k_{\parallel}$, the constancy of local energy flux, $v_k^2/\tau_{nl}$, yields $v_k \propto k_{\bot}^{-1/4}$. Consequently, the energy power spectrum is given by

\begin{equation}\label{BScaling}
E_{Bol}(k_\perp)\sim k_\perp^{-3/2}.
\end{equation}

GS95 and Boldyrev's models in fact predict similar diffusion of particles and magnetic field lines in spite of the fact that they have some strong physical differences (Jafari et al. 2018). Boldyrev's ideas have been criticized by Lazarian et al. (2015) whose results confirm Goldreich-Sridhar scaling. \\

\subsection{General Scheme of Stochastic Reconnection}\label{s4}

Resistivity is responsible for only one type of diffusion process (normal diffusion). A mere enhancement of the diffusion coefficient, e.g., by appealing to some anomalous diffusivity, does not change the nature of diffusion. Depending on the conditions present, the reconnection may rely on other diffusion processes as well. If we attribute a "wrong" diffusion process to the reconnection, our predicted rate would be unrealistic of course. For instance, in the inertial range of a turbulent cascade, the field lines diffuse away according to the Richardson diffusion which is a super-diffusion (mean square separation of field lines grows super-linearly with time). Overlooking turbulence and applying the normal diffusion caused by resistivity (in which mean square separation grows linearly with time) leads to the Sweet-Parker model. Thus the fact that the Sweet-Parker model does not work in highly turbulent astrophysical environments is not surprising. On the other hand, identifying the underlying diffusion process is one thing and modeling the diffusion in a realistic way is another. After modeling the field line diffusion, the reconnection rate may be estimated using, e.g., the conservation of mass in the current sheet. The large scale geometry of the reconnection zone may also affect the reconnection speed. As far as the scaling laws drive us, we may take care of the geometry by inserting a "geometrical factor" in the reconnection speed.

\iffalse Faraday's notion of force lines remains as an important visualization tool utilized even across the reconnection zones in resistive and turbulent fluids. Even in the presence of fully developed turbulence, its footprints are so obvious in the models of stochastic reconnection. However, the fuzzy field topology in a turbulent medium, where flux-freezing assumption breaks apart and the field diffuses away, is not suggestive of a picture in which the field lines persist in preserving their individual robust identity. The reconnection problem in astrophysics (reconnection in length scales much larger than those important in small scale physics of the current sheet) can be approached applying the diffusion laws for the magnetic field without appealing to the individual field lines (Eyink et al. 2013). This approach still applies the notion of field lines in describing their rms separation. However, it is done applying a statistical averaging with no emphasis on the individual lines.  \fi

\begin{figure}
\begin{centering}
\includegraphics[scale=.35]{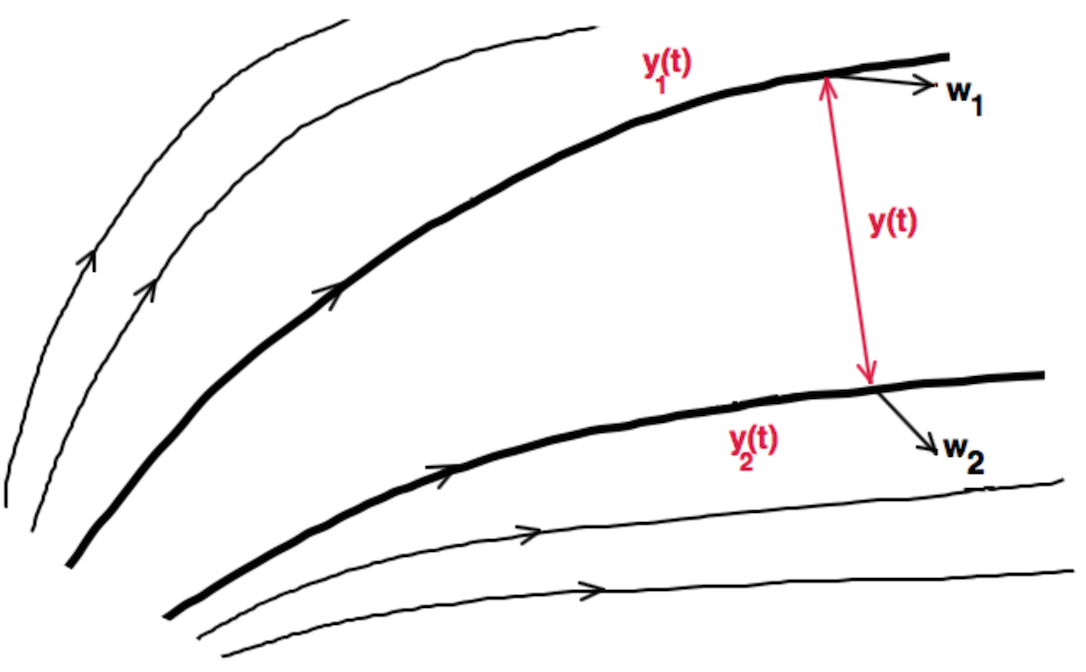}
\centering
\caption {\footnotesize {We can define a "relative motion" between two magnetic field lines in a resistive medium. Ensemble averaging over many field lines, the relative velocity $w=\langle|{\bf{w}}_1-{\bf{w}}_2|^2\rangle^{1/2}$ can be expressed in terms of the RMS separation, $y(t)=\langle |{\bf{y}}_1(t)-{\bf{y}}_2(t)|^2\rangle^{1/2}$, as $w=\dot y$.}}
\end{centering}
\label{Rel-motion-B}
\end{figure}

Fig.(\ref{Rel-motion-B}) illustrates a bundle of field lines highlighting two of them whose rms separation changes over time as

\begin{equation}\notag
w={d \over dt}\langle |{\bf{y}}_1(t)-{\bf{y}}_2(t)|^2\rangle^{1/2}.
\end{equation}
We can simplify our notation a little bit by denoting the r.m.s separation as $y(t)=\langle |{\bf{y}}_1(t)-{\bf{y}}_2(t)|^2\rangle^{1/2}$. With a fluid with resistivity $\eta$ we write

\begin{equation}\label{diff1}
{dy(t)\over dt}=\Big({\delta y\over \delta t} \Big)_{res}\simeq {y(t)\over \tau_R}\simeq {\eta\over y(t)}.
\end{equation}

The solution of this simple differential equation is $y(t)\simeq \sqrt{\eta t}$. This indicates a normal diffusion which is usually described using a random walk approach. In a random walk with the mean step size $\overline y=|\bf{y}_i|$ and time scale $\tau$, the average total distance vanishes $\langle {\bf{y}}\rangle=\langle \sum_i {\bf{y}}_i \rangle=0$, however, the rms distance does not; $\sqrt{\langle y^2\rangle}=\langle \sum_i y_i^2 \rangle^{1/2}=\sqrt{N \overline y^2}$. The number of steps is, of course, the total time divided by the average time for each step, $N=t/\tau$. This leads to the familiar result for normal diffusion with a linear growth $\langle y^2\rangle\sim D t$ where $D$ is the diffusion coefficient. (A non-linear growth corresponds to anomalous diffusion, $\langle y^2\rangle\sim D t^\alpha$, with super-diffusion for $\alpha>1$ and sub-diffusion for $\alpha<1$.) In an analogous manner, we may argue, the mean square distance that a particular magnetic field line diffuses by resistivity is given by

  \begin{equation}\label{15+1}
y^2(t) \sim \eta t.
 \end{equation}

In terms of scaling laws, the above result is understandable from the induction equation; $\partial_t{\bf{B}}=\eta\nabla^2\bf{B}$. As an aside note that in contrast ambipolar diffusion, which leads to a parallel electric field of the form $\eta_{am} \bf{J}$, preserves the field topology (Brandenburg \& Zweibel 1994) and therefore is not related to the present discussion.

\iffalse

Similarly, we may think of hyper-resistivity (originating from an electric field given by $\eta_H\nabla^2 \bf{J}$; see eq.(\ref{hyper-resistivity})) giving rise to a sub-diffusion;
  \begin{equation}\label{15+1}
y^4(t) \sim \eta_H t.
 \end{equation}
 
 \fi

Turbulence too has a diffusive effect on the magnetic field lines. The effect of the turbulent eddy motions on the magnetic field lines can be represented by an additional "RMS relative speed" between any pair of field lines. in analogy with eq.(\ref{diff1}), we can write
\begin{equation}\label{diff2}
{dy(t)\over dt}=\Big({\delta y\over \delta t} \Big)_{turb}\simeq {y(t)\over \tau_T}.
\end{equation}
To estimate $\tau_T$, take a turbulent fluid with negligible resistivity as an example. A simple dimensional analysis, in the inertial range of the turbulent cascade, leads to the eddy turnover time of order $\tau_T\sim \epsilon^{-1/3}y^{2/3}$ (Kolmogorov scaling). The solution is the Richardson diffusion
\begin{equation}
y^2(t)\sim \epsilon t^3,
\end{equation}
 which is of course a super-diffusion.

One can put all this into a compact formalism. The total "r.m.s speed" between any arbitrary pair of field line would be the sum of all "velocity contributions" as

\begin{equation}\label{diffusion-lag2}
\dot y={dy(t)\over dt}=\Big( {\delta y\over \delta t} \Big)_{turb}+\Big( {\delta y\over \delta t} \Big)_{res}+...=\sum_{i} \Big( {\delta y(t)\over \delta t}\Big)_i,
\end{equation}
where "..." indicates other diffusion mechanisms might be present as well. The above equation is the equation of motion and can be obtained from the following Lagrangian

\begin{equation}\label{diffusion-lag1}
{\cal{L}}={\dot y^2\over 2}+{1\over 2} \Big[ \sum_{i} \Big( {\delta y(t)\over \delta t}\Big)_i \Big]^2.
\end{equation}

The Euler-Lagrange equation for the above Lagrangian is a second order differential equation. It can be integrated though to obtain the first order equation given by eq.(\ref{diffusion-lag2}) provided that we set one initial condition as $\dot y(0)=0$. Then eq.(\ref{diffusion-lag2}) is solved with the second initial condition; $y(0)=y_0$. A non-trivial solution $y(t)$, valid even for the initial condition $y(0)=y_0\rightarrow 0$, would then mean that the field lines start or end at the same point---contradicted with a deterministic dynamics. Such a solution can only appear in a turbulent regime, in which case, the diffusion equation does not satisfy the mathematical requirements for the existence and uniqueness of the solutions. For example, neglecting all effects but turbulence with the time scale $\tau_T =\epsilon^{-1/3} y^{2/3}$ in the inertial regime, the equation of motion is found as $\dot y \sim y^{1/3}$ with "non-deterministic" solution $y^2\sim t^3$ valid even with $y(0)=0$. So the paradox is resolved simply by noting that there is no uniqueness theorem for a differential equation of the form $\dot y=A y^h$ for $h<1$ (Eyink et al. 2013).

In practice, we may estimate the terms inside the brackets in eq.(\ref{diffusion-lag1}) simply by appealing to the relevant time scales, hence the above treatment of diffusion problem looks like an effective field theory. For example, the turbulent velocity is estimated as $(\delta y/\delta t)_{turb}\sim y/\tau_T$ with eddy turnover time $\tau_T$ and similarly for resistivity $(\delta y/\delta t)_{res}\sim y/\tau_R$ with the time scale $\tau_R\sim y^2/\eta$. This term would be negligible in most astrophysical contexts because of the large magnetic Reynolds numbers; $y/\tau_R\sim \eta/y\ll 1$. We need to consider any $y$-dependence of $\tau_T$ by writing $\tau_T=\sum_n \alpha_n y^{\beta_n}$ with coefficients $\alpha_n$ and $\beta_n$ that may depend on viscosity and the nature of turbulence such as the energy transfer rate $\epsilon$, energy injection scale $l$, etc. On any scale, only one term in this sum will be large and dominate the others. For a resistive and turbulent fluid, eq.(\ref{diffusion-lag2}) becomes

\begin{equation}\label{general-diffusion}
{dy\over dt}=y\Big({1\over \tau_R}+{1\over  \sum_n \alpha_n y^{\beta_n} } \Big).
\end{equation}
Let us divide up the turbulence time scale as $\sum \alpha_ny^{\beta_n}=\tau_h+\tau_y+\tau_d$. To estimate the hydrodynamic turbulent time scale $\tau_h$, we first note that the characteristic length contraction of a particular field line between adjacent eddies is $\Delta x=l_{\bot}^2/l_{\parallel}$, $l_{\bot}$ and $l_{\parallel}$ as discussed above. The corresponding time scale for the diffusion of the field lines would be $\tau_h\sim(l_\parallel/l_\perp^2)y^2/V_A$ where $V_A$ is the local Alfv\'en speed. If the rms distance happens to be in the dissipative regime, where the time scale depends only on the viscosity $\nu$ and energy transfer rate $\epsilon$, the dissipative time scale is given by $\tau_d\sim\lambda_{\parallel, d}/V_A$ where $\lambda_{\parallel, d}$ is given by eq.( \ref{parlambda}): $\lambda_{\parallel,d} \simeq  l^{1/2}(V_TV_A)^{-1/4}({V_A/ V_T})^{4/3}\nu^{1/2}$. Therefore, $\tau_d\simeq V_A^{-1/2}l^{1/2}(V_TV_A)^{-1/4}(V_A/V_T)^{4/3}\nu^{1/2}$. In the inertial range, we have $\tau_y\sim \lambda_\parallel/V_A$ where $\lambda_\parallel$ is given by GS95 scaling given by eq.(\ref{67}) with $y\equiv k_\perp^{-1}$. Also, we estimate the resistive time scale as $\tau_R\sim y^2/\eta$. Putting all this together, we arrive at

\begin{equation} \label{general}
{dy\over dx}\simeq {\eta/V_A\over y}+\frac{y}{ay^2+b y^{2/3}+c\sqrt{\nu}} .
\end{equation}
where $a=(l_\parallel/l_\perp^2) $, $b=l^{1/3}(V_A/V_T)^{2/3}$ and $c= l^{1/2}(V_TV_A)^{-1/4}(V_A/V_T)^{3/4}$. We have also used $dx=V_A dt$. Eq.(\ref{general}) represents the diffusion equation of the magnetic field lines in a turbulent medium with arbitrary viscosity and resistivity.

In a reconnection zone of length $\sim L_x$, one can use the mass conservation to estimate the reconnection speed $V_R$ in terms of the the outflow speed $V_{eject}$:
\begin{equation}
V_R\simeq \Lambda_g \Big({y(t)\over t} \Big)_{t=L_x/V_{eject}},
\end{equation}
where $\Lambda_g$ is a dimensionless constant which depends on the large scale geometry of the current sheet. In most cases, the outflow or ejection velocity would be of order the local Alfv\'en speed $V_{eject}\simeq V_A$.

The general formalism developed above leads to the well-known stochastic reconnection models for small magnetic Prandtl numbers $Pr_m\lesssim 1$ (see e.g., LV99; Lazarian et al. 2004; Lazarian et al. 2012; Eyink et al. 2013) as well as large magnetic Prandtl numbers $Pr_m\gg 1$ (Jafari et al. 2018). In a non-viscous and non-turbulent environment, the reconnection follows the Sweet-Parker scaling, $V_R\sim \eta^{1/2}$. This can be seen using eq.(\ref{general}) by ignoring the effects of turbulence and viscosity but not resistivity;

\begin{equation} \notag
\frac{d y^2}{dx}\simeq \frac{\eta}{V_A} \rightarrow V_R\simeq {V_A\over S^{1/2}}.
\end{equation}

But, in fact, it is resistivity that can be neglected in most astrophysical media which are highly vulnerable to development of turbulence. Furthermore, as discussed before, even in a quiet environment, the current sheet instabilities will eventually produce turbulence. In plasmas with resistivity less than or at most of order viscosity, the reconnection becomes independent of the resistivity:
\begin{equation} \notag
\frac{d y^2 }{dx}\simeq {y^2 \over b y^{2/3}}\rightarrow V_R\lesssim V_T \Gamma,
\end{equation}

where $\Gamma$ is a geometric factor which depends on whether the current sheet is smaller, $L_x<l$, or larger, $L_x>l$, than the large eddies (Kowal et al. 2012). This is the stochastic reconnection of Lazarian and Vishniac (LV99). The stochastic reconnection speed given above was obtained by Lazarian et al. (2012) by using Richardson law. This corresponds to $n=1/3$ in 
\begin{equation}\label{85-1}
\frac{dy}{dt}=\delta u_y=\alpha y^n,
\end{equation}
with the solution $y^2=[y_0^{-(n-1)}-\alpha (n-1) t]^{-2/{(n-1)}}$ where $0<n<1$, and $\alpha$ is a constant. Therefore, we get the turbulent energy spectrum as $E(k) \propto k^{-(2n+1)}$ (Eyink et al. 2013). 

In the following sections, we review stochastic reconnection in high $\beta$ plasmas for special but important cases of moderate and large magnetic Prandtl numbers $Pr_m=\nu/\eta$.\\

\begin{figure}
\includegraphics[scale=.25]{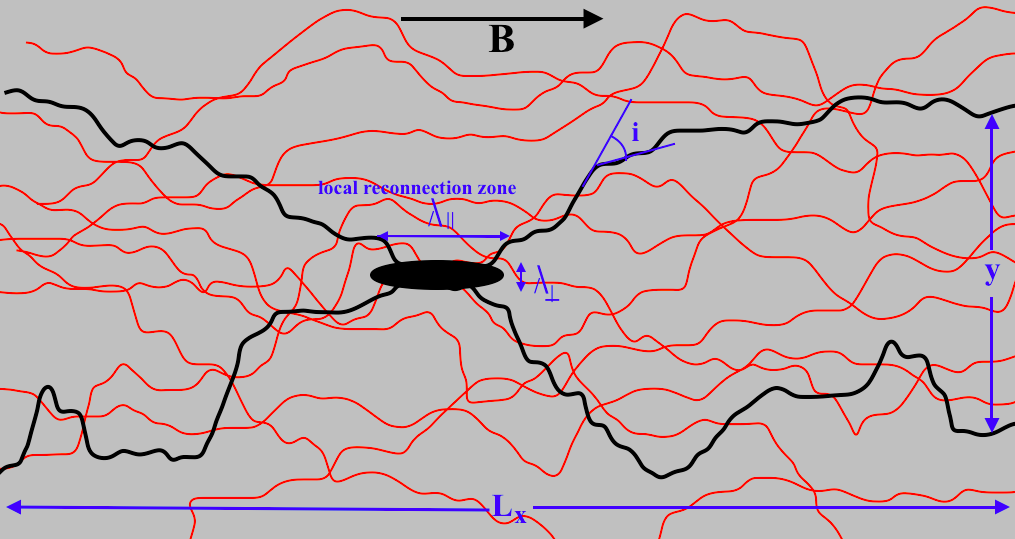}
\centering
\caption {\footnotesize {An exaggerated view of the stochastic wandering of the turbulent magnetic field components as discussed by Lazarian and Vishniac (LV99). Assuming that the angle $i$, indicating the field line deviation from its initial direction, is small for sufficiently small separation $y$, the vertical rms separation $\langle y^2\rangle^{1/2}$ can be estimated by writing $i \simeq \langle y^2\rangle^{1/2}/\lambda_\parallel$ where $\lambda_\parallel$ is the wavelength parallel to the mean field $\bf{B}$ and corresponds to the vertical wavelength $\lambda_\perp=\langle y^2\rangle^{1/2}$; see eq.(\ref{LV-equation}). In a turbulent medium, however, the magnetic field lines are unlikely to possess a robust identity and geometrical arguments of LV99 should be replaced with a more physical picture.  }}
\label{stochastic}
\end{figure}

\subsection{ Stochastic Reconnection For Moderate $Pr_m$}\label{ss50}

The stochastic model proposed by Lazarian and Vishniac (LV99) is based on the notion of stochastic magnetic field line wandering for moderate magnetic Prandtl numbers, $Pr_m\sim 1$. Unlike the conventional Sweet-Parker model, in stochastic reconnection, it is assumed that there are many independent and simultaneous local reconnection events occurring over the large scale current sheet; see Fig.(\ref{ST1}). These local small scale Sweet-Parker reconnections, with the speed $V_{R,loc}$, occur only over a scale $\lambda_\parallel \ll L_x$ (and not over the  current sheet length $L_x$). The parallel length scale $\lambda_\parallel$ is the scale over which a field line deviates from its initial direction by the thickness of the Ohmic dissipation layer; $\lambda_\perp\simeq \eta/V_{R,loc}$ (LV99). 

One constraint on the achievable reconnection rates comes from the matter outflow from the reconnection layer. This leads to an upper limit on the global reconnection rate. To see this, consider a bundle of parallel magnetic field lines filling a region of width $y$ at some particular point. The field lines will spread out perpendicular to the direction of the mean magnetic field. LV99 estimated the field diffusion by appealing to the following diffusion equation: 

\begin{equation}\label{LV-equation}
{d \langle y^2 \rangle\over dx}\simeq {\langle y^2 \rangle\over \lambda_\parallel},
\end{equation}
where the parallel scale $\lambda_\parallel$, corresponding to the perpendicular rms distance $\lambda_\perp=\sqrt{\langle y^2\rangle}$, is given by eq.(\ref{67}). In the original arguments of LV99, there is no physical explanation for the origin of the above equation and apparently it emerges out of a simple geometrical consideration at the eddy scale; see Fig.(\ref{stochastic}). However, we can see that this equation is a special case of eq.(\ref{general}) in which viscosity, resistivity and the normal diffusion of field lines are assumed to be of negligible roles. For small $\langle y^2 \rangle $, integrating the above expression leads to 
\begin{equation}\label{77}
\langle y^2 \rangle^{1/2}\simeq \frac{x^{3/2}}{l_\parallel^{1/2}} \left(\frac{V_T}{V_A}\right).
\end{equation}

Thus, the thickness of the outflow layer increases with the parallel displacement $L_x$ as
\begin{equation}\label{78}
\langle y^2 \rangle^{1/2}\sim
L_x \left(\frac{L_x}{l_\parallel}\right)^{1/2} \left(\frac{V_T}{V_A}\right); l_\parallel>L_x,
\end{equation}
and
\begin{equation}\label{78+1}
\langle y^2 \rangle^{1/2}\sim
\left({L_x l_\parallel}\right)^{1/2} \left(\frac{V_T}{V_A}\right); l_\parallel<L_x.
\end{equation}

LV99 obtained the following upper limit on the reconnection speed using $V_{LV} \sim V_A \langle y^2 \rangle ^{1/2}/L_x$;

\begin{equation}\label{79}
V_{LV}\leq V_T \; \textit{min} \; \left\{ \left (\frac{L_x}{l_\parallel} \right)^{1/2}, \left (\frac{l_\parallel}{L_x} \right)^{1/2}\right\}.
\end{equation}

This upper limit is obviously independent of resistivity which makes it fast. Therefore, unlike the Sweet-Parker model in which the microphysical diffusivity limits the outflow, in the stochastic model of LV99, the large scale magnetic field wandering determines the width of the outflow region. Two physical mechanisms are responsible in building up the large reconnection speed of LV99. First, only a tiny fraction of magnetic field lines would experience the resistive diffusion making the reconnection rate independent of resistivity, which is negligible in astrophysical systems. Second, a large number of field lines enter to the reconnection zone simultaneously (Kowal et al. 2009). When the turbulent effects become extremely weak, the range of field line wandering becomes much smaller than the width predicted by the Sweet-Parker layer, $L_x/\sqrt{S}$ and consequently the two models become indistinguishable. 

\iffalse
\begin{figure}
\includegraphics[scale=.42]{stochastic1}
\centering
\caption {\footnotesize {Stochastic reconnection as suggested by Lazarian and Vishniac (LV99). Richardson diffusion can bring a pair of oppositely oriented field lines into a close contact so they can reconnect. This is a local Sweet-Parker type reconnection, in this picture, occurring near the origin. After the reconnection, the field lines wander away stochastically. In fact, many independent but simultaneous local reconnections of this type occur everywhere in the large scale reconnection region. The separation $\langle y^2\rangle^{1/2}$ is defined as the rms distance between a pair of field lines as a function of $x$, the distance from the initial point where local reconnection occurs, along the $x$ axis; $\langle y^2\rangle^{1/2}=\langle y^2\rangle^{1/2}(x)$.  }}
\label{stochastic1}
\end{figure} 
\fi

Yet, there is another condition imposed by the local mass conservation. In order for matter to escape from the reconnection layer, it should travel a distance $\lambda_\parallel(\lambda_\perp)$, with $\lambda_\perp=\delta$ as the width of the local reconnection layer, along the local magnetic field. This requires 
\begin{equation}\label{new1}
V_{R,loc} \lambda_\parallel \simeq V_A \lambda_\perp.
\end{equation}
Similar to the Sweet-Parker model, at this fundamental level, we assume that the Ohmic diffusion derives the reconnection. Consequently,
\begin{equation}\label{new2}
V_{R,loc}\simeq {\eta\over \lambda_\perp}.
\end{equation}
Recalling that $\lambda_\parallel=k_\parallel^{-1}$ and $\lambda_\perp=k_\perp^{-1}$, we can use eq.(\ref{67}) along with equations (\ref{new1}) and (\ref{new2}) to find the local reconnection rate as
\begin{equation}
V_{R,loc}\simeq (V_T V_A)^{1/2}\Big({\eta\over l_\parallel V_A} \Big)^{1/4}.
\end{equation}

One implication here is that the largest expected wave-number in the turbulent cascade is given by
\begin{equation}
k_\perp\simeq {V_{R,loc}\over \eta}\simeq {(V_TV_A)^{1/2} \over l_\parallel V_A}\Big( {V_A l_\parallel\over \eta}\Big)^{3/4}.
\end{equation}

This wave-number is also the perpendicular wave-number of the field fluctuations that control the local reconnection events. However, we are interested in the global reconnection rate. One may argue that along the length of the reconnection zone $L_x$, there are $L_x/\lambda_\parallel$ local reconnection layers. Therefore,
\begin{equation}
V_{R, global}\simeq k_\parallel L_x V_{R, loc}\simeq V_A  {L_x\over l_\parallel} \Big( {V_A l_\parallel\over \eta} \Big)^{1/4} \Big( {(V_TV_A )^{1/2}\over V_A} \Big)^3.
\end{equation}
This condition is generally less restrictive than eq.(\ref{79}). Therefore the matter flow from the reconnection zone does not limit the reconnection speed.

In hydrodynamic turbulence, small scale diffusion and large scale shear together give rise to Richardson diffusion where the mean square separation between two particles grows as $t^3$ when the rms separation exceeds the viscous damping scale. In MHD turbulence, there seems to be a similar mechanism where at later times the separation is independent of the microscopic transport coefficients. Note that the latter derivation of the LV99 reconnection speed may look trivial, however, the fundamental reasoning behind that should be justified using the recent understanding of frozen-in magnetic field lines for turbulent MHD plasmas (see Lazarian et al. 2012 and references therein). Another point worth mentioning is that even though we used ensemble averaging in the above approach, LV99 made no use of averaging over turbulent ensembles, and in fact the only use they made of ensembles was to get a measure of the typical wandering distance $\delta$ of the magnetic field lines. The advantage of this new derivation is that it avoids complex considerations of the cascade of reconnection events discussed in LV99 (Kowal et al. 2012).\\

\subsection{Stochastic Reconnection For Large $Pr_m$}\label{ssVSR}

What is the effect of a viscosity much larger than the diffusivity on the stochastic reconnection---large magnetic Prandtl number regime, $Pr_m=\nu/\eta>1$? Both analytical (see e.g., Lazarian et al. 2004; Jafari et al. 2018) and numerical (Cho et al. 2002; Cho et al. 2003; Schekochihin et al. 2004) work point to the important role that a large viscosity plays in MHD turbulence. Here we discuss the regime of large magnetic Prandtl number.

Let us begin with a general picture in which the medium is resistive, viscous and turbulent. The reconnection region would be filled with a small scale field component $\bf{b}$ wandering stochastically along the the direction of a mean field $\bf{B}$; see Fig.(\ref{Stochastic-diffusion}). In MHD turbulence, there are two dissipation scales; the first being the viscous damping scale, $l_{d}$ while the second one is the magnetic diffusion scale, $l_{m}$. In fully ionized collisionless plasmas, both of these dissipation scales are very small. In partially ionized plasmas, they can be very different but always we have $l_{d}\gg l_{m}$. When the energy reaches the viscous damping scale, $l_{d}$, the kinetic energy would be dissipated, similar to hydrodynamic turbulence, but at this scale magnetic energy does not dissipate. Hydrodynamic cascade will stop at viscous damping scale, $l_{d} \sim \nu^{3/4}\epsilon^{-1/4}$ that is

\begin{equation}
\label{89}
l_{d} \sim l _\perp\left ( {\nu\over l_\perp V_T}\right)^{3/4},
\end{equation}

where we have used the sub-Alfv\'enic energy transfer rate $\epsilon \sim v_l^4/(l_\parallel V_A)\simeq V_T^2V_A/l_\parallel$ ($v_l=\sqrt{V_TV_A}$) with critical balance condition (GS95; LV99; Jafari et al. 2018) $V_T l_\parallel\simeq l_\perp V_A$. Similarly, the magnetic damping scale is given by

\begin{equation}
\label{90}
l_{m} \sim l_\perp \left ( {\eta\over l_\perp V_T}\right)^{3/4}.
\end{equation}

At the viscous damping scale $l_{d}$, the hydrodynamic cascade stops due to viscosity, however, magnetic field would still be sheared (Lazarian et al. 2004; Jafari et al. 2018). The width of the  reconnection layers, created by the viscosity damped turbulence, can be determined using the eddy size at the viscous damping scale $l_{d}$. Consequently, the width of such a reconnection layer would be roughly $k_{\bot,d}^{-1}=l_{d}=\lambda_{\perp,d}$. The corresponding parallel length at the damping scale, $\lambda_{\parallel, \; d}$, can be obtained by substituting $k_{\bot}^{-1}=l_{d}$ in eq.(\ref{67});

\begin{equation}\label{parlambda}
\lambda_{\parallel,d} \simeq  l_\parallel^{1/2}(V_TV_A)^{-1/4}\Big({V_A\over V_T}\Big)^{3/4}\nu^{1/2}.
\end{equation}

  \begin{figure}
\includegraphics[scale=.4]{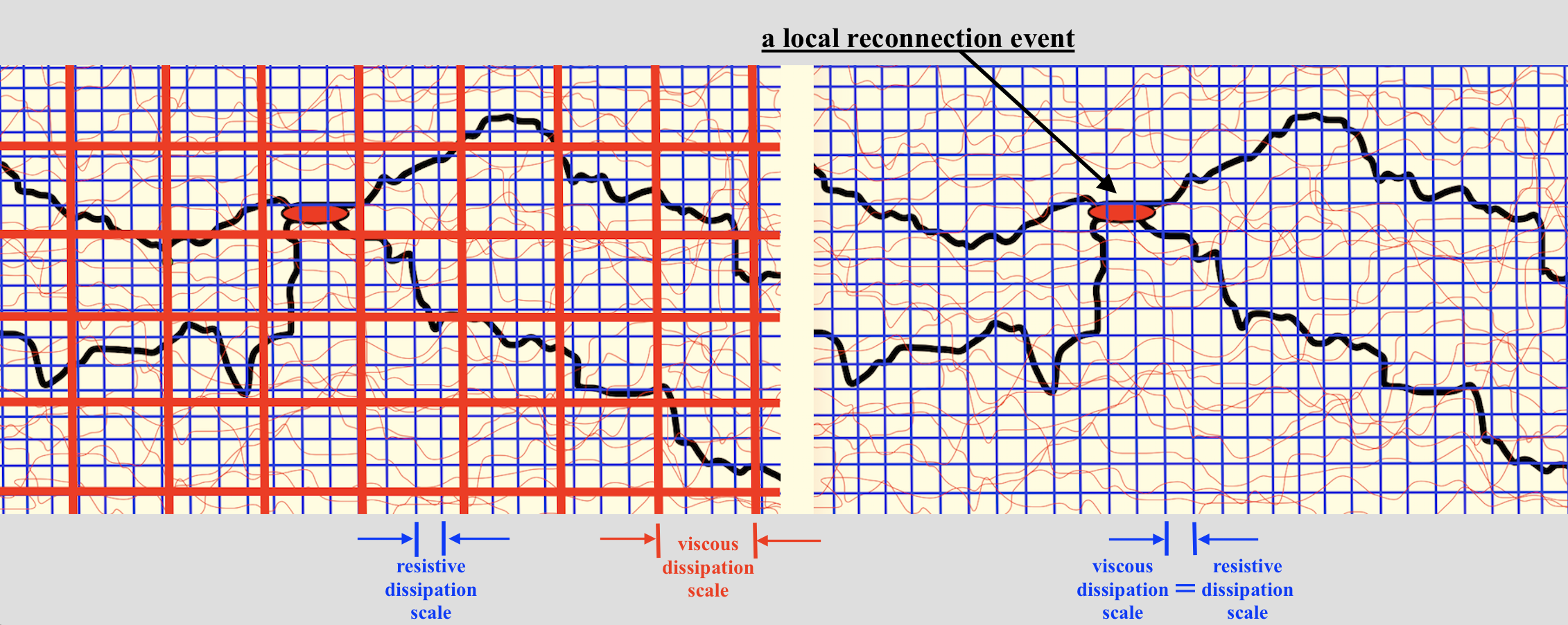}
\centering
\caption {\footnotesize {Schematics of stochastic reconnection in two noisy environments with different magnetic Prandtl numbers. In each picture only one pair of field lines and their local reconnection zone (red ellipse) are highlighted. Right: Stochastic reconnection with $Pr_m=\nu/\eta \sim 1$. Each blue cell represents the resistive length scale $l_m\sim \eta^{3/4}$ which is the same as the viscous length scale $l_d\sim\nu^{3/4}$. This is stochastic reconnection with magnetic Prandtl number of order unity discussed by Lazarian and Vishniac (LV99). Note that the width of the local reconnection zone is of order the dissipation length $l_d\sim l_m$ (side of the blue cells). Left: Stochastic reconnection with viscosity larger than diffusivity; $Pr_m\sim 5.3$. Each red cell has a length of order the viscous length scale $l_d$ which is larger than the resistive length scale $l_d\sim 3.5 l_m$ (side of the blue cells). In this regime, reconnection occurs below the viscous damping scale and the effect of viscosity may be important. }}
\label{viscous}
\end{figure} 

 Below $l_{d}$, the energy cascade rate $\tau_{cas}^{-1}$ is scale independent, and thus the constant magnetic energy rate $b_l^2 \tau_{cas}^{-1}$ leads to $b_l\sim const.$;
\begin{equation}\label{power1}
E_b(k)\sim k^{-1}.
\end{equation}
The above expression is the magnetic power spectrum by the assumption that the curvature of field lines is almost constant $k_\parallel\sim const.$. This is consistent with a cascade driven by repeated shearing at the same large scale. Also, the component of the wave-vector parallel to the perturbed field does not change much and therefore any change in $k$ comes from the third direction (Jafari et al. 2018).
 
To find the kinetic spectrum, we define the filling factor $\phi_l$ as the fraction of the volume that contains strong field perturbations with a scale $k^{-1}$ (Lazarian, Vishniac \& Cho 2003). Inside these volumes, we denote the velocity and magnetic fields with a hat such that $v_l^2=\phi_l \hat v_l^2$ and $b_l^2=\phi_l \hat b_l^2$. The viscous force $\nu \nabla^2 \bf{v}$ is balanced with the magnetic tension $\bf{j}\times \bf{B}$;
\begin{equation}\label{LVC10}
{\nu\over l^2}\hat v_l\sim max \; \Big(\hat b_l k_c, B_0 k_{\parallel,c} \Big)\hat b_l,
\end{equation}
where $k_c=l^{-1}_{d}$ with a corresponding parallel wave-vector component $k_{\parallel, c}$. Since $\hat b_l\geq b_l$ and we can use the critical balance $B_0 k_{\parallel, c}\sim b_l  k_c$ to write eq.(\ref{LVC10}) as
\begin{equation}\label{LVC11}
{\nu\over l^2}\hat v_l\sim B_0 k_{\parallel, c}\sim b_l^2 k_c.
\end{equation}
Below the viscous damping scale, $l_{dv}$, the motions are sheared continuously at a rate $\tau_s^{-1}$. The equilibrium is reached when they generate a comparable shear;
\begin{equation}\label{LVC12}
{\hat v_l \over l}\sim \tau_s^{-1}\sim constant.
\end{equation}
We combine this with eq.(\ref{LVC11}) to get $(k_c l) \hat b_l^2\sim const.$ that, comparing with the definition $b_l^2=\phi_l \hat b_l^2$ gives us $\phi_l\sim k_c l$. Thus, assuming a constant filling factor, $\phi_l\sim const.$, we get the kinetic spectrum $E_v(k)\sim k^{-5}$. Otherwise, we have
\begin{equation}
E_v(k)\sim k^{-4}.
\end{equation}

\begin{figure} 
\centering
\includegraphics[scale=.15]{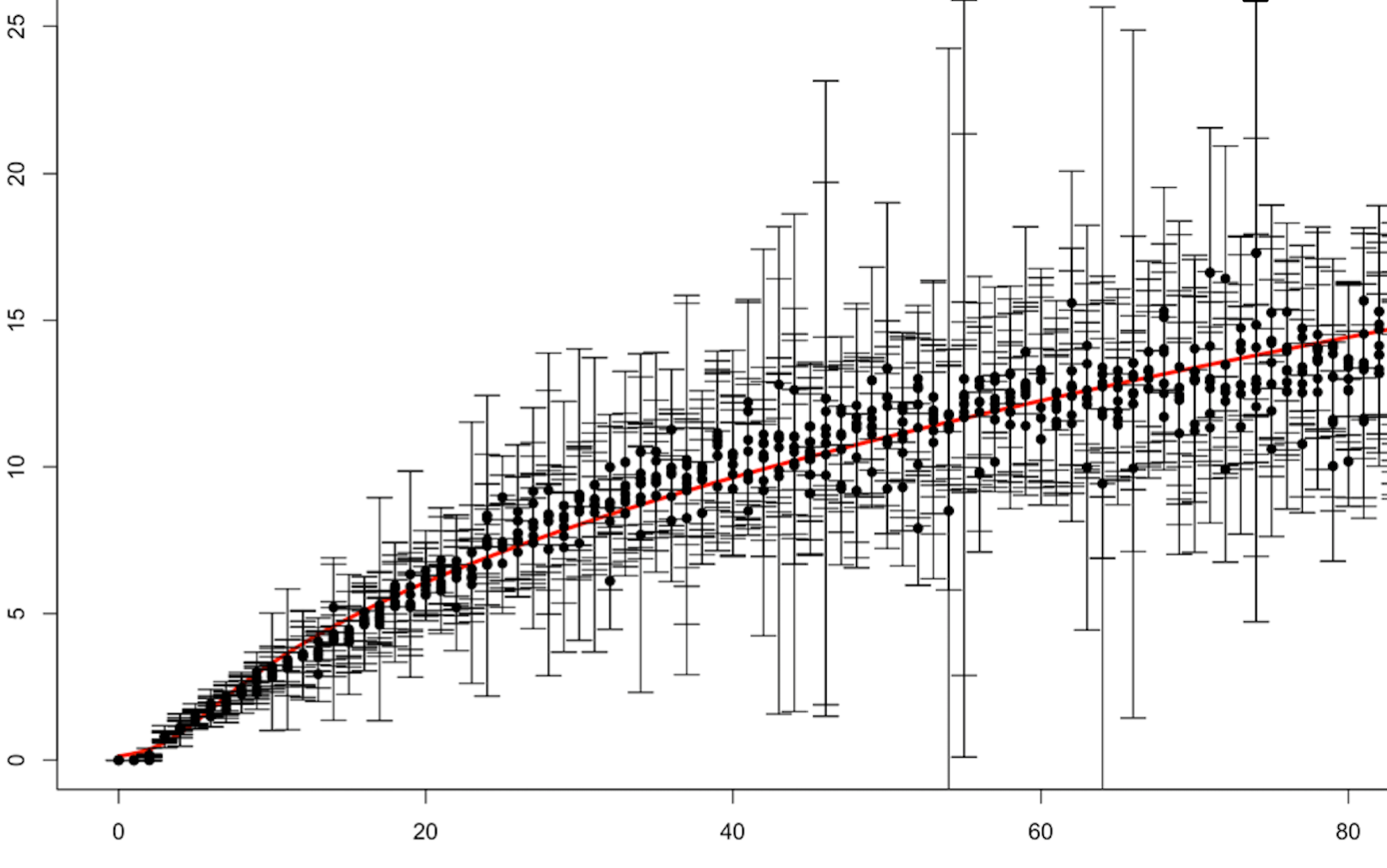}
\caption {\footnotesize {Separation of magnetic field lines of Kowal et al. (2012) compared with the analytical solution of eq.(\ref{general}) presented by the red line. The magnetic Prandtl number is of order $10$ (Jafari et al. 2018). }}
\label{Separation}
\end{figure}

To find the reconnection speed in large magnetic Prandtl number regime, Jafari et al. (2018) used eq.(\ref{general}) with modified coefficients $a=a_1 l_\parallel/l_\perp^2$, $b=b_1 l_\parallel^{1/3}(V_A/V_T)^{2/3}$ and $c= c_1 l_\parallel^{1/2}(V_TV_A)^{-1/4}(V_A/V_T)^{3/4}$. The numerical factors $a_1 = 1.32$, $b_1 = 0.36$ and $c_1 = 0.41$ are constants introduced to match the solution to the available numerical data obtained by Kowal et al. (2012) in which the injection power $P=V_T^2V_A/l_\parallel$ and the Alfv\'en speed $V_A$ are set to unity and the injection wavelength $k_{inj}=l_\parallel^{-1}=8$. For these simulations, numerically, we have $a=a_1/4$ and $b=b_1$ and $c=c_1$. Fig.(\ref{Separation}) shows a comparison of the solution of eq.(\ref{general}) with the numerical data obtained by Kowal et al. (2012). As mentioned before, we combine the solution $y=y(t)$  of eq.(\ref{general}) with the conservation of mass; $V_R=V_{eject} (\langle y^2 \rangle)^{1/2}/L_x$ where $V_{eject}$ is the ejection speed of plasma out of the reconnection zone. 

In the regime of low viscosity, one may take $V_{eject}$ to be roughly equal to the Alfv\'en speed $V_A$ although it's actually a large fraction of it. In the high viscosity regime, $ \nu /\eta \gg 1$, however, the viscosity should have some effect on the outflow if the viscous timescale is shorter than the outflow timescale, $\langle y^2 \rangle /\nu  < L_x/V_{eject}$. One can rewrite this condition using $\langle y^2\rangle/l_\perp^2 \sim L_x/l_\parallel $, $V_A l_\perp\simeq l_\parallel V_T$ and defining the Reynolds number as $Re=l_\perp V_T/\nu$ where $V_T$ is the velocity characteristic of largest strongly turbulent eddies. We find
\begin{equation}\label{c_1}
Re< {V_A\over V_{eject}}.
\end{equation}

This condition is obviously not to be satisfied in nature or numerical simulations. Consequently, viscosity can hardly affect the outflow speed (Jafari et al. 2018). Nevertheless, viscosity can affect the reconnection rate through changing the outflow channel width. Viscosity may affect reconnection either by hindering the material outflow from the reconnection zone or by narrowing the channel width of the reconnection zone. If the perpendicular separation is totally in the inertial cascade, viscosity would not play an important role. In other words, if $y> l_{d}$ even for $x=L_x$, viscosity has negligible effect. In the viscous regime, the diffusion equation reads
\begin{equation}
\frac{d y^2 }{dx}\simeq \frac{y^2}{\lambda_{\parallel,d}} +\frac{\eta}{V_A},\;\;y(0)=0,
\end{equation}
with the solution
\begin{equation}
\langle y^2 \rangle \simeq \Big({\eta\over V_A}  \Big)\lambda_{\parallel,d} \exp{\Big({L_x\over \lambda_{\parallel, d}} \Big)}.
\end{equation}
The condition $y \lesssim l_{d}$ then is written as
\begin{equation}
{L_x\over l_\parallel} \Big({l_\perp V_A\over \nu}\Big)^{1/2} <1+ \ln \Big( {\nu\over \eta}\Big).
\end{equation}
This condition can be put in a more compact form in terms of $Re=l_\perp V_T/\nu$ and $Pr_m=\nu/ \eta$;
\begin{equation}
 Re^{1/2} <\Big({ l_\parallel\over L_x} \Big) \Big[1+\ln ( Pr_m)\Big].
 \label{limit}
\end{equation}

The width of the outflow region, $y=\langle y^2\rangle^{1/2}$, is therefore unaffected unless the magnetic Prandtl number is exponentially larger than the Reynolds number.

\begin{figure}
\begin{centering}
\includegraphics[scale=.6]{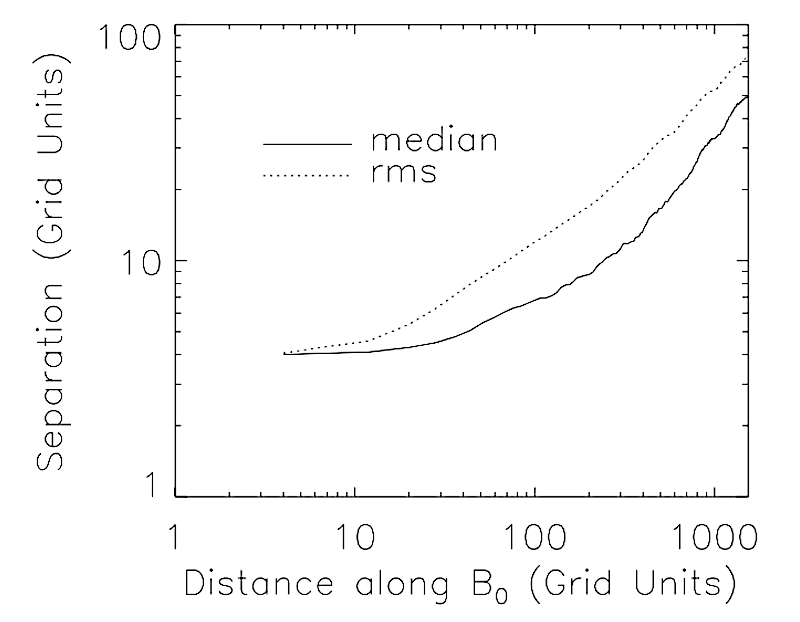}
\caption {\footnotesize {The rms and median separation of magnetic field lines plotted versus the distance parallel to the mean field in the viscosity damped turbulence using a $384^3$ grid by Lazarian, Vishniac and Cho (2004). The median tracks the mean at a slightly smaller amplitude consistent with the exponential growth. This exponential growth in the dissipative regime can also be seen looking at the analytic result given by eq.(\ref{viscosity-damped}). }}
\end{centering}
\label{LVC(2004)}
\end{figure}

For large $Pr_m$, we need to take into account the terminating effect of the viscosity on the turbulent cascade. The general expectation is that viscosity would slightly slow down the reconnection if it is much larger than the resistivity. Fig.(\ref{Separation}) compares the general solution of eq.(\ref{general}) with the numerical simulations performed by Kowal et al. (2009; 2012). In the latter regime, the hydrodynamic turbulent cascade stops at the viscous damping scale, $l_{d}\sim \nu^{3/4}$ but the magnetic structures are not suppressed at this length scale because the magnetic damping scale is much smaller, $l_m\sim \eta^{3/4}\ll l_{d}$. Below the damping scale, we find 
\begin{equation} \label{viscosity-damped}
\frac{dy^2}{dx}\simeq {\eta\over V_A}+{ y^2 \over \lambda_{\parallel,d}}\rightarrow y^2\sim {\eta\over V_A}\lambda_{\parallel,d}\Big[ \exp{(x/\lambda_{\parallel,d})}-1\Big],
\end{equation}
which, for $x\gg \lambda_{\parallel,d}$, becomes Lyapunov exponential growth whose exponent is inversely proportional to the square root of the viscosity. Fig.(\ref{LVC(2004)}) illustrates this exponential growth of field line separation in a set of numerical simulations preformed by Lazarian et al. (2004). As an aside also note that ignoring resistivity in the diffusion and assuming that it leads to an initial separation we get the initial condition $y^2(0)=\eta\lambda_{\parallel,d}/V_A$ instead of $y(0)=0$ which leads to a Lyapunov growth. At some parallel distance $x=L_{RR}$, known as Rechester-Rosenbluth distance, the perpendicular separation reaches the dissipation scale $\lambda_{\perp,d}$. We find
\begin{equation} \label{viscosity-damped}
L_{RR}\simeq \lambda_{\parallel,d} \ln{\Big({\lambda_{\perp,d}\over y_{in}} \Big) },
\end{equation}
where we have defined an initial vertical separation as $y_{in}=(\eta \lambda_{\parallel,d}/V_A)^{1/2}$ (Lazarian et al. 2015).

The other possible effect of viscosity on reconnection is through altering the local small scale reconnection events. We already know that the condition that viscosity affect the large scale outflow is a very stringent condition, which seems to indicate that viscosity will almost never be important.  So the other question is whether or not the small scale effect of viscosity can be significant.

Yet, another effect viscosity can have on reconnection arises from topologically distinct local reconnections at small scales. In stochastic reconnection, the global reconnection speed is in fact the outcome of many reconnection events happening simultaneously at small scales. As put by LV99, this sets an upper limit on the global reconnection speed estimated as the small scale eddy reconnection speed times the number of independent eddies along a field line;
\begin{equation}
V_{rec}< N_{eddy} v_{rec,eddy},
\end{equation}
where $v_{rec,eddy}$ is the reconnection speed within an individual eddy. For a magnetic Prandtl number of unity, the regime LV99 is intended for, the reconnection speed in an undamped eddy is always the local turbulent velocity. In this case, the number of eddies along a field line scales with $k_{\|}$. Therefore, this scale-dependent limit scales as $k_\perp^{1/3}$ and rises continuously to the dissipation scale. It is always less important than global limits on the reconnection speed. However, for a magnetic Prandtl number greater than unity, this limit may fall to interesting values at small scales and play an important dynamic role (see, e.g., Lazarian et al. 2004). 

Here we follow Jafari et al. (2018) to construct a simple model that takes this effect into account.  To begin with, we note that this model requires revised estimates of $v_{rec,eddy}$ and $N_{eddy}$, both of which will be affected by viscosity. Velocities are strongly suppressed below the viscous dissipation scale, therefore, we first consider reconnection speeds at the dissipation scale and then show that the reconnection speed constraint arising from that scale is the most stringent. As previously shown (Cho et al. 2002), for large magnetic Prandtl numbers we expect the magnetic field perturbations to show a highly intermittent distribution on scales below the Kolmogorov scale with the spontaneous appearance of thin current sheets. The width of these current sheets is given by $\delta\simeq \sqrt{\eta\tau_d}$ where $\tau_d$ is the coherence time for magnetic structures on the damping scale. The associated local reconnection speed is $v_{rec, diss}\simeq \sqrt{\eta/\tau_d}$.  Reconnection is inhibited on these scales, so the reconnection rate will be less than the eddy turnover rate. This implies that turbulence creates magnetic structures, or "knots", that resolve more slowly than the eddy turn over time on larger scales. Suppose the reconnection speed is limited by the constraint set at the Kolmogorov scale. The persistence time for large scale magnetic structures is then given by
\begin{equation}
\tau_p^{-1}\simeq {k_\perp\over k_{\perp,diss}}\tau_d^{-1}{k_{\|,diss}\over k_{\|}}\propto k_\perp^{1/3}.
\end{equation}
The magnetic perturbations on the affected scales will no longer track the velocity perturbations, rather it will scale as
\begin{equation}
{\rho v_k^2\over\tau_c}\simeq {b_k^2\over \tau_p}.
\end{equation}
Thus the magnetic fluctuations on a scale $k_\perp^{-1}$ will scale as $k_\perp^{-1/6}$, instead of the Goldreich-Sridhar estimate of $k_\perp^{-1/3}$. Thus the effect of a magnetic Prandtl number greater than one is to create a magnetic fluctuation spectrum which is relatively shallow and a persistence time for magnetic structures which is relatively long. In order to estimate $\tau_d$, we note 

\begin{equation}
\tau_d^{-1}\simeq k_{\perp,diss} \left({\eta\over\tau_d}\right)^{1/2} \left({\tau_d\over\tau_{c,diss}}\right)^{1/2},
\end{equation}
where we have dropped the logarithmic correction to $N_{eddy}$ in this expression since it is unlikely to be significant. We rewrite the above expression as
\begin{equation}
{\tau_{c,diss}\over\tau_d}\simeq \left({\eta\over\nu}\right)^{1/2}.
\end{equation}
The local reconnection speed at the Kolmogorov scale, therefore, is given by the local eddy speed divided by $\sqrt{Pr_m}$.
 
As for the global reconnection speed, we note that the constraint imposed by the viscous suppression of small scale motions implies the following condition:
 \begin{equation}
V_{rec} < v_{rec,eddy} \left({\eta\over\nu}\right)^{1/2} N_{eddy}\simeq V_T { Re^{1/4} Pr_m^{-1/2}\over 1+\ln(Pr_m)}{L_x\over l_\|}.
\label{plimit}
 \end{equation}

So for $\nu=\eta$ this is not an interesting limit, but for $Re$ not very large and resistivity appreciably less than viscosity it can be the controlling limit on reconnection (and not the outflow zone width).  In particular the dividing line between viscosity dominated reconnection and unimpeded reconnection is no longer overwhelmingly in favor of unimpeded reconnection. Therefore, the critical magnetic Prandtl number does not scale exponentially with the Reynolds number but close to its square root. The simple model outlined above, however, is based on the assumption of a large degree of intermittency below the Kolmogorov scale. This may not be very realistic. In addition, numerical simulations of the viscose regime have currently low resolutions and their results are affected by a damping scale very close to the eddy scale. The small Reynolds numbers accessible by numerical simulations at present may be the reason why they hardly show any effect associated with viscosity (Jafari et al. 2018).

 \begin{figure}
\includegraphics[scale=.85]{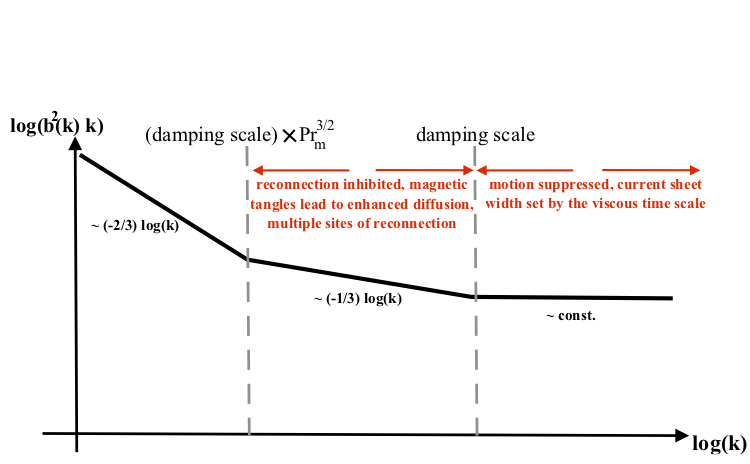}
\centering
\caption {\footnotesize {Magnetic fluctuation spectrum $b^2(k)k$: Below viscous damping scale, the hydrodynamic motions are suppressed and we have a flat spectrum for magnetic fluctuations. The width of the current sheet is controlled by viscosity in this regime. Above the viscous damping scale, and below the scale set by $(damping\;scale)\times Pr_m^{3/2}$, reconnection is inhibited while diffusion is enhanced by magnetic tangles.}}
\label{x}
\end{figure} 

The magnetic structures are sheared below the viscous damping scale with a scale independent energy cascade rate and we find a power spectrum for the magnetic field as $E_b(k)\sim k^{-1}$: eq.(\ref{power1}). It follows that more power concentrates on the small scales than predicted by the Goldreich-Sridhar spectrum (GS95). With a viscosity of order, or slightly larger than the resistivity, the  the outflow width remains independent of the small scale physics, which is similar to the LV99 model. Even with viscosity appreciably larger than resistivity, the width of the outflow region is unaffected unless the magnetic Prandtl number is exponentially larger than the Reynolds number. The viscosity is even less important when the current sheet instabilities dominate over external turbulence. For significantly large Prandtl numbers, i.e., of order $\sqrt{Re}$, the magnetic field perturbations cannot relax and a flat magnetic power spectrum extends below the viscous damping scale; see Fig.(\ref{x}).

\section{Summary and Discussion}\label{s6}

Our focus in this review paper was on the astrophysical reconnection in which the current sheet width is large enough that small scale physics becomes irrelevant e.g., scales much larger than the ion inertial length. This is a special but extremely important case of magnetic reconnection. Since 1960s, different reconnection schemes have been proposed that rely on a broad range of different physical mechanisms such as Bohm diffusion, Hall effect and turbulence. \iffalse Yet, many of these models

(i) they presume a very idealistic environment not usually present in real systems (e.g., the Sweet-Parker model),\\
(ii) they rely on a potentially unstable configuration (e.g., the Petschek model),\\
(iii) their requirements are not usually met in real reconnection events and, even if they are, their effect on astrophysical reconnection is probably minor (e.g., Bohm diffusion and Hall effect). \fi

The "failure" of a model in predicting a realistic reconnection rate can be the result of some flaws in the underlying physical model. However, it may also be due to the "wrong" application of a model which is not suitable to the conditions present. For instance, the Sweet-Parker model is constructed for, and can be quite successful in, a highly conductive inviscid plasma with a stable laminar flow. Of course, these conditions are not present in real situations specially in astrophysics. Any boundary between oppositely oriented magnetic fields would be unstable to instabilities making the medium turbulence, let alone the external sources of turbulence e.g., supernovae. The resulting reconnection in such a noisy environment would also enhance the turbulence. Thus, in some sense, it is the incorrect application, rather than the incorrect model, that leads to disagreement with observations. In fact, many different phenomena become involved in the reconnection simultaneously, of course, some in major and some in minor ways. All these can contribute to reconnection through diffusing the magnetic field lines in the fluid. Based on this idea, we developed a mathematical formalism for astrophysical reconnection in which, depending on the time scales present, one of the involved diffusion processes would dominate over the others resulting in the fastest reconnection realizable in the environment. One caveat, to bear in mind, is that this is only a book-keeping device which uses the known concepts of diffusion to quantify reconnection in astrophysical systems: it does not constitute a new reconnection model.

The simple resistive MHD description based on a single fluid picture (Parker 1957; Sweet 1958; Petschek 1964) fails in predicting a reasonable reconnection rate. The assumption that the electrons and ions move together in thin current sheets is so naive. In these reconnection zones, the ions are demagnetized and the relative drift velocity of the ions and electrons is typically very large (Yamada et al. 2010). Current sheet thicknesses encountered in the magnetopause or similar environments are of order the ion inertial length (Vasyliunas, 1975; Yamada et al. 2010) while those encountered in astrophysics are much larger. Thus, in an astrophysical context, neither the resistive MHD nor its modifications with the Hall effect or other similar treatments would work in predicting a realistic reconnection picture. An overlooked factor is turbulence which is ubiquitous in astrophysical media but needs a special treatment based on the notion of field line stochasticity and diffusion.

Although some authors have argued that turbulence may slow down the reconnection rate (e.g., Kim \& Diamond 2001), the general expectation is that turbulence enhances reconnection. Some authors have proposed a reconnection-enhancing role for turbulence appealing to hyper-resistivity or turbulent diffusivity (e.g., Guo et al. 2012). However, most of these models are based on loose theoretical grounds and are highly controversial. In fact, turbulence and current sheet instabilities are probably the most important elements affecting the astrophysical reconnection events. Reconnection itself generates, or at least enhances, the turbulence which in turn enhances the reconnection rate. So, the dominant diffusion process in astrophysical reconnection would be the Richardson diffusion slightly modified by viscosity (at very small scales) and possibly, but not so probably, some other mechanisms. Magnetic diffusivity, which is almost always negligible, and viscosity, which is usually small but can be much larger than the diffusivity, both have minor effects but might be important in developing instabilities and also in very small scales so should be included for self-consistency. 

In an ideally non-turbulent environment with zero viscosity, magnetic reconnection follows the Sweet-Parker scheme, $V_R\sim \eta^{1/2}$. This is confirmed by numerical simulations but it is not realistic in two ways. First, there is almost no such a "quiet" environment in astrophysics. Second, even an initially quiet environment would become turbulent because of instabilities or as a result of reconnection itself. In turbulent plasmas with a magnetic Prandtl number of order unity or smaller magnetic reconnection is stochastic with no dependence on neither resistivity nor viscosity. The stochastic model, proposed by LV99, has proved successful in turbulent collisional plasmas and collisionless plasmas with small neutral fractions. In plasmas with very high magnetic Prandtl numbers, e.g., partially ionized plasmas, the role of viscosity may become slightly important but its overall effect seems to be small. At very high magnetic Prandtl numbers, the reconnection process is affected by turbulent cascade and probably viscosity but not resistivity. 

The general picture in the viscous regime is as follows. The hydrodynamic cascade is interrupted at the viscous damping scale, $l_{d}\sim \nu^{3/4}$, however, the magnetic structures are not suppressed below this scale since the resistive damping scale is smaller than viscous damping scale, $l_{d}\gg l_m\propto \eta^{3/4}$. The magnetic perturbations get sheared below the damping scale with a scale independent magnetic energy cascade rate. The magnetic power spectrum is given by $E_b(k)\sim k^{-1}$ with much more power concentrated on the small scales if compared with the celebrated MHD turbulence model of Goldreich-Sridhar. In a medium with the viscosity of order or slightly larger than the resistivity, the outflow thickness is independent of the small scale physics which resembles the LV99 scheme. With viscosity larger than resistivity, the thickness of the outflow zone remains unaffected unless the magnetic Prandtl number is exponentially larger than the Reynolds number. The viscosity has a less important role when the current sheet instabilities, generating internal turbulence, dominate over the external turbulence.

The important conclusions can be summarized as follows:

(i) Reconnection proceeds in the presence of turbulence, no matter generated internally or externally. Reconnection enhances turbulence as turbulence enhances reconnection.

(ii) Small scale physics, such as considerations of enhanced resistivities, would become irrelevant in noisy astrophysical reconnection zones. In real turbulent systems, stochastic reconnection can be understood in terms of diffusion of field lines whose rms separation, $y(t)$, has an effective Lagrangian in terms of the involved diffusion time scales $(\delta y/\delta t)_i$; ${\cal{L}}\rightarrow {\dot y^2/ 2}+ \Big[  \sum_i\Big( {\delta y(t)/ \delta t}\Big)_i \Big]^2/2$.

(iii) The reconnection rate in typical astrophysical systems is independent of resistivity. For $Pr_m\sim 1$, reconnection rate is also independent of viscosity. For $Pr_m>1$, the width of the outflow region remains unaffected unless the magnetic Prandtl number is exponentially larger than the Reynolds number. The outflow velocity remains unaffected too unless the Reynolds number is of order unity. Thus, viscosity seems to have negligible effect on both the outflow width and ejection velocity. This would imply that in typical astrophysical systems, viscosity remains unimportant. However, for $Pr_m>1$, if the Reynolds number is not too large while the magnetic Prandtl number is, viscosity may slow down the global reconnection through its effects on the small scale reconnection events. This leads to a threshold for the magnetic Prandtl numbers larger than the square root of the Reynolds number.  This effect is also expected to flatten the magnetic fluctuation power spectrum.

\iffalse
 \begin{figure}
\includegraphics[scale=.7]{Ali2014}
\centering
\caption {\footnotesize {The growth rates of reconnected magnetic flux in the explosive growth phase versus magnetic Prandtl number, $Pr$, for three different resistivities. The solid lines mark the power scaling for $Pr\gg 1$ (Ali et al. 2014).}}
\label{Ali2014}
\end{figure} 

\fi

\section{References}
\raggedright

Baty H., Priest, E. R., \& Forbes, T. G. 2006, Phys. Plasmas 13, 022312

Becker, U., Neukirch, T., \& Schindler, K. 2001, J. Geophys. Res. 106, 3811

Beresnyak, S. 2012, MNRAS, 422, 4
 
Beresnyak, S. 2013, arXiv:1301.7424 [astro-ph.SR]

Beresnyak, S. 2016, ApJ, 834, 1

Biskamp, D. 1984, Phys. Lett. A, 105, 124

Biskamp, D. 1986, Phys. Fluids, 29, 1520

Biskamp, D. 1993, Nonlinear Magnetohydrodynamics, CUP

Biskamp, D. 1996, Ap\&SS, 242, 165

Biskamp, D. 2000, Magnetic Reconnection in Plasmas, Cambridge Monogr. on Plasma Phys., Cambridge Univ. Press, New York

Biskamp, D. 2003, Magnetohydrodynamic Turbulence, pp. 310. ISBN 0521810116. Cambridge, UK: Cambridge University Press

Boffetta, B.,  Ecke. R. E. 2012, Annu. Rev. Fluid Mech. 44, 427

Boldyrev, S. 2005, ApJ, 626, 1, L37

Boldyrev, S. 2006, Phys. Rev. Lett, 96, 11

Brandenburg, A., \& Zweibel, E. G. 1994, ApJ, 427, L91

Brandenburg, A., \& Zweibel, E. G. 1995, ApJ, 448, 734

Brown, M., 1999, Phys. Plasmas 6, 1717

Browning, P., \& Lazarian, A. 2013, Space Sci. Rev., 178, 325

Cho, J. \& Vishniac, E. 2000, ApJ, 539, 273

Cho, J. ,Lazarian, A., \& Vishniac, E. T. 2002,ApJ, 564, 291

Cho, J., Lazarian, A., \& Vishniac, E. T. 2003, Turbulence and Magnetic Fields in Astro-
physics, 614, 56

Cothran, C. D., Landreman, M., Brown, M. R., \& Matthaeus, W. H. 2003, Geophys. Res. Lett. 30, 1213

Cranmer, S. R., Asgari-Targhi, M., Miralles, M.P., Raymond, J.C., Strachan, L., Tian, H., \& Woolsey L. N. 2015, Phil. Trans. R. Soc. A 373, 20140148

Diamond, P. H., \& Craddock, G. G. 1990, Comments on the Plasma Physics of Controlled Fusion, 13, 287

Drake, J. F., Swisdak, M., Cattell, C., et al. 2003, Science, 299, 873

Drake, J. F., Swisdak, M., Che, H., \& Shay, M. A. 2006, Nature, 443, 553

Drake, J. F. 1995, in Physics of the Magnetopause, edited by 
Song, P., Sonnerup, B., \& Thomsen, M., AGU Monograph Vol.
90  AGU, Washington, D.C. , p. 81

Eyink, G. L., 2011, Phys. Rev. E 83, 056405

Eyink, G. L., Lazarian, A., \& Vishniac, E. T., 2011, ApJ, 743, 1

Eyink, G., Vishniac, E., Lalescu, C., Aluie, H., Kanov, K., Burger, K., Burns, R., Meneveau, C., Szalay, A. 2013, Nature, 497, 7450, 466

Forbes, T. G. 2007, in Reconnection of Magnetic Fields: Magnetohydrodynamics and Collisionless Theory and Observations, edited by J. Birn and E. R. Priest  Cambridge University Press, Cambridge , p. 16.

Frisch, U. 1995, Turbulence: The Legacy of A.N. Kolmogorov, Cambridge University Press 

Furth, H.P., Killeen, J., \& Rosenbluth, M.N. 1963, Phys. Fluids 6, 459

Goldreich, P., \& Sridhar, S. 1995, ApJ, 438, 763 (GS95)

Goldreich, P. 1997, ApJ, 485, 680

Guo, Z.B., Diamond, P.H. \& Wang, X.G. 2012, ApJ, 757, 173

Haugen, N. E. L., Brandenburg, A., \& Dobler, W. 2005, Phys.
Rev. E 70, 016308

Hayashi, T., \& Sato, T. 1978, J. Geophys. Res. 83, 217.

Huang, Y. M., \& Bhattacharjee, A. 2010, Phys. Plasmas, 17, 062104

Huang, Y., Bhattacharjee, A., \& Forbes, T. G. 2013, Phys. Plasmas, 20, 8

Innes, D. E. , Inhester, B, Axford, W. I., \& Wilhelm, K. 1997, Nature, 386, 811

Iroshinkov, P. S. 1963, AZh, 40, 742

Jafari, A., Vishniac, E. T., Kowal, G., \& Lazarian, A. 2018, ApJ, 860, 52

Jemella, B. D., Drake, J. F.  \& Shay, M. A. 2004, Phys. Plasmas 11, 5668

Karimabadi, H., Krauss-Varban, D., Huba, J.D., \& Vu, H.X. 2004. J. Geophys. Res. 109, A09205

Kim, E-j. \& Diamond, P. H. 2001, ApJ, 556, 1052

Kivelson, M., \& Russell, C. 1995, Introduction to Space Physics  Cambridge University Press, London 

Kolmogorov, A. 1941, Doklady Akademiia Nauk SSSR, 30, 301

Kowal, G., Lazarian, A., Vishniac, E. T., \& Otmianowska-Mazur, K. 2009, ApJ, 700, 63

Kowal, G., Lazarian, A., Vishniac, E. T., \& Otmianowska-Mazur, K. 2012, Nonlin. Processes Geophys., 19, 297

Kraichnan, R. H. 1965, Phys. Fluids, 8. 1385

Kraichnan, R. H., Montgomery, D. 1980,  Rep. Prog. Phys., 43

Krall, N., \& Liewer, P. 1971, Phys. Rev. A 4, 2094

Krause, F., \& R{\"a}dler, K.-H. 1980, Mean-Field Magnetohydrodynamics and Dynamo Theory, Akademie-Verlag, Berlin

Kulsrud, R. M. 2000, Earth Planets Space, 53, 417

Kulsrud, R.M. 2005, Plasma Physics for Astrophysics. Princeton, NJ, Princeton Univ. Press

Kuznetsova, M. M., Hesse, M, Rastätter, L., Taktakishvili, A., Toth, G., De Zeeuw, D. L. , Ridley, A., \& Gombosi, T. I. 2007, J. Geophys. Res. 112, A10210

Lazarian, A., \& Vishniac, E. T. 1998, arXiv:astro-ph/9804166v1

Lazarian, A., \& Vishniac, E. T. 1999, ApJ, 517, 700 (LV99)

Lazarian, A., Petrosian, V., Yan, H., \& Cho, J. 2003, Proceedings of Beaming and Jets in Gamma Ray Bursts, 45

Lazarian, A., Vishniac, E. T., \& Cho, J. 2004, ApJ, 603, 180

Lazarian, A., Eyink, G. L., \& Vishniac, E. T. 2012, Phys. Plasmas, 19, 012105

Lazarian, A., Eyink, G. L., Vishniac, E. T., \& Kowal, G. 2015, Phil. Trans. R. Soc. A 373: 20140144

Loureiro, N. F., \& Uzdensky, D. A. 2016, Plasma Physics and Controlled Fusion, 58, 1

Lovelace, R. V. E. 1976, Nature, 262, 649

Lyutikov, M., \& Lazarian, A. 2013, Space Sci. Rev., 178, 459

Malyshkin, L. M., Linde, T., \& Kulsrud, R. M. 2005, Phys. Plasmas, 12, 102902

Matthaeus, W.H., \& Lamkin, S.L. 1985. Phys. Fluids 28:303

Matthaeus, W.H., \& Lamkin S. L. 1986. Phys. Fluids 29:2513

Matthaeus, W.H., Wan, M., Servidio, S., Greco, A., Osman, K. T., Oughton, S., \& Dmitruk, P. 2015, Phil. Trans. R. Soc. A 373, 20140154

McBride, J., Ott, E., Boris, J., \& Orens, J. 1972, Phys. Fluids
15, 2367

Mestel, L. 1985, MNRAS, 212, 275 

Milano, L. J., Matthaeus, W. H., Dmitruk, P., \& Mont-
gomery, D. C. 2001, Phys. Plasmas,  8, 2673.

Moffatt, H. K. 1978, Magnetic field generation in electrically conducting fluids, Cambridge

M{\"u}ller, W.-C., Biskamp, D., \& Grappin, R. 2003, Phys. Rev.
E 67, 066302

M{\"u}ller, W.-C., \& Grappin, R., Phys. Rev. Lett. 2005, 95, 114502.

Ono, Y., Morita, A., Katsurai, M., \& Yamada, M. 1993, Phys. Fluids B 5, 3691

Papadopoulos, K. 1977, Reviews of Geophysics and Space Physics, 15, 1, 113

Parker, E. N. 1957, J. Geophys. Res., 62, 509

Parker, E. N. 1970, ApJ 162, 665

Parker, E. N. 1979, Cosmological Magnetic Fields (Oxford, Clarendon)

Parker, E. N. 1993, ApJ, 408, 707

Petschek, H. E. 1964, The Physics of Solar Flares, AAS-NASA Symposium (NASA SP-50), ed.
W. H. Hess (Greenbelt, MD: NASA), 425

Priest, E., \& Forbes, T. 1986, J. Geophys. Res. 9, 5579

Priest, E., \& T. Forbes, 2000, Magnetic Reconnection: MHD
Theory and Applications,  Cambridge University Press, Cambridge 

Priest, E. \& Forbes, T. 2002, Astron. Astrophys. Rev., 10, 313

Priest, E. 1986, Mit. Astron. Ges. 65, 41

Sato, T., \& Hayashi, T. 1979, Phys. Fluids 22, 1189.

Samtaney, R., Loureiro, N., Uzdensky, D., Schekochihin, A., \& Cowley, S.
2009, Phys. Rev. Lett., 103, 105004

Schekochihin, A. A., Cowley, S. C., Maron, J. L., \& McWilliams, J. C. 2004, ApJ, 612, 276

Scholer M. 1989. J. Geophys. Res. 94, 8805

Shay, M. A., Drake, J. F., Rogers, B. F., \& Denton, R. E. 2001, J. Geophys. Res. 106, A4, 3759

Shibata, K., \& Tanuma, S. 2001, Earth Planets Space, 53, 473

Smith, D., Ghosh, S., Dmitruk, P., \& Matthaeus, W. H. 2004, Geophys. Res. Lett. 31, L02805

Sonnerup, B. U. \"{O}. 1970, J. Plasma Phys. 4, 161

Strauss, H. R. 1988, ApJ, 326, 412

Sturrock, P. A. 1966, Nature, 211, 695

Sweet, P. A. 1958, in IAU Symp. 6, Electromagnetic Phenomena in Cosmical Plasma, ed. B.
Lehnert (New York: Cambridge Univ. Press), 123

Ugai, M., \& Tsuda, T. 1977, J. Plasma Phys. 17, 337

Uzdensky, D., \& Kulsrud, R. 2000, Phys. Plasmas 7, 4018

Uzdensky, D., Forest , C., Ji, H., Townsend, R., \& Yamada, M. 2009, arXiv:0902.3596v1 

Vasyliunas, V. 1975, Rev. Geophys. Space Phys. 13, 303

Vishniac, E. T. 1995a, ApJ, 446, 724

Vishniac, E. T. 1995b, ApJ, 451, 816

Waelbroeck, F. L. 1989, Phys. Fluids B 1, 2372

Wang, X., Bhattacharjee, A., \& Ma, Z. W. 2001, Physical Review Lett., 87, 265003

Wang, X., Ma, Z. W., \& Bhattacharjee, A. 1996, , Phys. Plasmas, 3(5), 2129

Wang, Y., Kulsrud, R., \& Ji, H. 2008, Phys. Plasmas 15,
122105

Wilmot-Smith, A. L., Priest, E. R., \& Horing, G. 2005, Geophysical and Astrophysical fluid dynamics, 99, 177

Yamada, M, Kulsrud, R, \& Ji, H. 2010, Reviews of Modern Physics, 82, 603

Yamada, M., Yoo, J., \& Myers, C. E. 2016, Physics of Plasmas 23, 055402

Zweibel, E. G., \& Brandenburg, A. 1997, ApJ, 478, 563

Zweibel, E. G. 2002, ApJ, 567, 962

Zweibel, E. L., \& Yamada, M. 2009, Annu. Rev. Astron. Astrophys., 47, 291

Zweibel, E. L., \& Yamada, M. 2016, Proc. R. Soc. A 472, 20160479

\end{document}